\documentclass[usenatbib,a4paper]{mn2e}

\usepackage{xspace}
\usepackage{amsmath}
\usepackage{epsfig}
\usepackage{defs}
\usepackage{longtable}
\usepackage{multirow}

\title[Testing Feedback-Modified Dark Matter Haloes]{Testing Feedback-Modified Dark Matter Haloes with Galaxy Rotation Curves: Estimation of Halo Parameters and Consistency with $\Lambda$CDM Scaling Relations}

\author[H. Katz et al. ]{Harley Katz$^{1}$\thanks{E-mail: hk380@ast.cam.ac.uk}, Federico Lelli$^{2}$, Stacy S. McGaugh$^{2}$, Arianna Di Cintio$^{3}$, 
\newauthor Chris B. Brook$^{4}$ and James M. Schombert$^{5}$\\
$^1$Institute of Astronomy and Kavli Institute for Cosmology, University of Cambridge, Madingly Road, Cambridge, CB3 0HA, UK\\
$^2$ Department of Astronomy, Case Western Reserve University, Cleveland OH, 44106, USA\\
$^3$Dark Cosmology Centre, Niels Bohr Institute, University of Copenhagen, Juliane Maries Vej 30, DK-2100 Copenhagen, Denmark\\
$^4$Ramon y Cajal Fellow, Departamento de F\'isica Te\'orica, Universidad Aut\'onoma de Madrid, 28049 Cantoblanco, Madrid, Spain\\
$^5$Department of Physics, University of Oregon, Eugene OR, 97403, USA\\
}

\begin{document}

\maketitle

\begin{abstract}
Cosmological $N$-body simulations predict dark matter (DM) haloes with steep central cusps (e.g. NFW, Navarro et al. 1996).  This contradicts observations of gas kinematics in low-mass galaxies that imply the existence of shallow DM cores.  Baryonic processes such as adiabatic contraction and gas outflows can, in principle, alter the initial DM density profile, yet their relative contributions to the halo transformation remain uncertain.  Recent high resolution, cosmological hydrodynamic simulations (Di Cintio et al. 2014, DC14) predict that inner density profiles depend systematically on the ratio of stellar to DM mass (M$_*$/M$_{\text{halo}}$).  Using a Markov Chain Monte Carlo approach, we test the NFW and the M$_*$/M$_{\text{halo}}$-dependent DC14 halo models against a sample of 147 galaxy rotation curves from the new {\it Spitzer} Photometry and Accurate Rotation Curves (SPARC) data set.  These galaxies all have extended H{\small I} rotation curves from radio interferometry as well as accurate stellar mass density profiles from near-infrared photometry.  The DC14 halo profile provides markedly better fits to the data compared to the NFW profile.  Unlike NFW, the DC14 halo parameters found in our rotation curve fits naturally fall within two standard deviations of the mass-concentration relation predicted by $\Lambda$CDM and the stellar mass-halo mass relation inferred from abundance matching with few outliers.  Halo profiles modified by baryonic processes are therefore more consistent with expectations from $\Lambda$ cold dark matter ($\Lambda$CDM) cosmology and provide better fits to galaxy rotation curves across a wide range of galaxy properties than do halo models that neglect baryonic physics.  Our results offer a solution to the decade long cusp-core discrepancy.
\end{abstract}


\section{Introduction}
Late-type galaxies (spirals and irregulars) possess large disks of cold gas (H{\small I}) that follow nearly circular orbits.  This gas can be used to trace the gravitational potential of a galaxy well beyond its stellar component where dark matter (DM) is expected to dominate \citep{Bosma1981,vanalbada1985}. Historically, H{\small I} rotation curves are fit by building a mass model for stars and gas based on observations and assuming a spherical DM halo with a given density profile \citep{vanalbada1985}. These density profiles are either empirically motivated or predicted from cosmological, DM-only simulations. The empirical DM profiles have inner, constant-density cores and provide good fits to the data over a broad mass range and for different assumptions of mass-to-light ratio, but have no basis in cosmology \citep{deBlok2008,Begeman1991,vanalbada1985}.  The density profiles from DM-only simulations, instead, have central cusps and provide poor fits to low-mass and low-surface-brightness (LSB) galaxies \citep{deBlok2001,deBlok2002,Gentile2004,KdN2006,KdN2008,KdN2009}.  These results are shown to be robust even when correcting for small scale non-circular motions that may be present in observed rotation curves \citep{Oh2008}. 

The density profiles of DM haloes can be affected by various baryonic processes.  First, DM haloes may adiabatically contract: baryons can pull more DM into the centre as the gas cools and condenses \citep{Blumenthal1986,Gnedin2004,Sellwood2005}.  This is of particular importance for high-mass galaxies; however, detailed studies show that not all high-mass galaxies are well fit by adiabatic contraction \citep{Katz2014}.  In terms of expansion, energetic feedback driven outflows \citep{Navarro1996,Read2005,Pontzen2012} and dynamical friction \citep{Elzant2001,weinberg2002,Johansson2009} can input enough energy into the halo in order to cause a significant decrease in central density and counteract the contraction. 

In many high resolution cosmological simulations, strong stellar feedback from massive stars and supernovae may drive outflows of large quantities of gas during repeated starburst events, causing an overall expansion of the  DM haloes and the formation of DM cores \citep{Navarro1996,Read2005,Governato2010,Pontzen2012,Schaye2015,Onorbe2015,Chan2015}.  Previous studies have shown that properties of simulated galaxies exposed to strong stellar feedback can be consistent with the inner density slopes observed in dwarf galaxies \citep{Oh2011}.  It has become clear that the core-formation process crucially depends on (M$_*$/M$_{\text{halo}}$) \citep{DC2014a,DC2014,Brook2015a,Brook2015b}, which relates the amount of energy from star formation to the gravitational potential energy of the halo. It is therefore necessary to test the predicted M$_*$/M$_{\text{halo}}$ dependency of core formation using a large sample of galaxies, spanning a broad range in stellar masses, rotation velocities, and gas fractions. 

In this paper, we compare two theoretical models for the density profiles of DM haloes: the NFW profile predicted from cosmological DM-only simulations \citep{Navarro1996}, and the DC14 model derived from cosmological galaxy formation simulations that include the effects of baryonic processes on their host DM haloes \citep{DC2014a,DC2014}.  Adopting the NFW model implicitly assumes that baryons have no effect on halo structure.  In contrast, the DC14 model exhibits a variable density profile that accounts for both cusps and cores depending on the ratio M$_*$/M$_{\text{halo}}$ that parameterizes the net effect baryons have in restructuring haloes.  

This paper is organized as follows. In Section~\ref{RCMM1}, we briefly describe the $Spitzer$ Photometry and Accurate Rotation Curves (SPARC) data set \citep{SPARC}.  In Section~\ref{RCfit}, we outline our Markov Chain Monte Carlo (MCMC) technique used to fit the galaxy rotation curves. In Section~\ref{results}, we present the results of the fitting the halo models to the galaxy rotation curves as well as how their parameters fit into the greater context of $\Lambda$ cold dark matter ($\Lambda$CDM).  Finally in Section~\ref{DandC} we illustrate how the DC14 profile provides a statistically significant improvement over the NFW model and discuss how this relieves the tension of the ``core-cusp" problem.  

\begin{figure*}
\centerline{\includegraphics[scale=0.5]{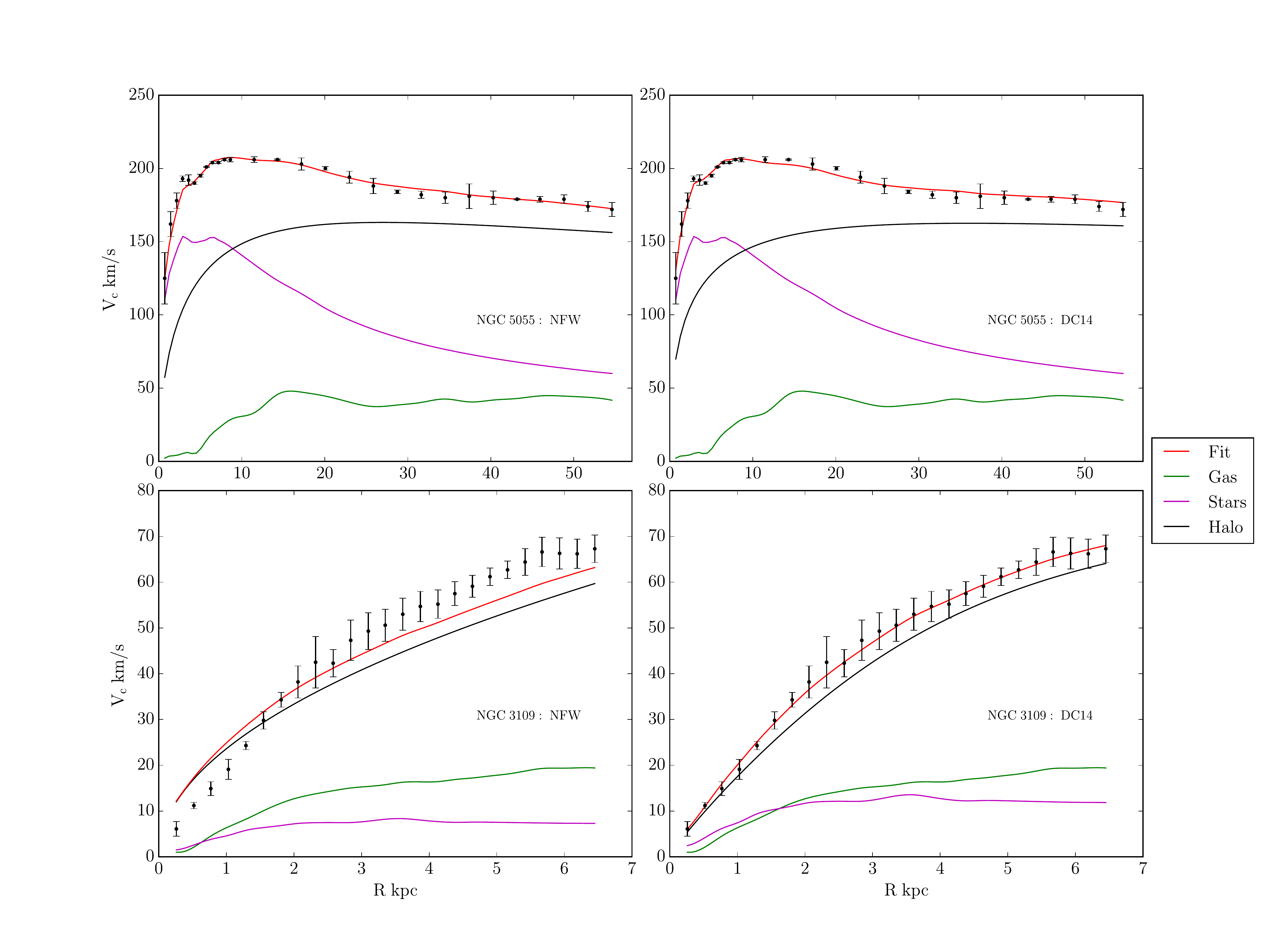}}
\caption{Maximum posterior fits for NFW haloes (left) and DC14 haloes (right) for a high (top) and a low-mass (bottom) galaxy.  The NFW and DC14 haloes differ little for high-mass and high-surface-brightness galaxies, giving comparable fits.  At low-mass and LSB, the NFW fit provides a poor description of the data.  This example illustrates the generic failure of cuspy halo models fit to data for LSB galaxies \protect{\citep{deBlok2001,KdN2009}}.  The baryon-modified DC14 halo provides a much better fit.}
\label{RCcomp}
\end{figure*}

\section{Rotation Curves and Mass Models}
\label{RCMM1}
This study is based on the SPARC data set \citep{SPARC}. Briefly, we collected more than 200 extended H{\small I} rotation curves from previous compilations \citep{Begeman1991, Sanders1996}, large surveys \citep{deBlok1996, Verheijen2001, Noordermeer2007, Swaters2009}, and several individual studies. These are the results of $\sim$30 yr of interferometric observations at 21 cm. The rotation curves were derived using similar techniques and software; hence, they form a relatively homogeneous data set. The specific technique used to derive the rotation curve depends on galaxy inclination ($i$) and data quality: (i) for galaxies with high-quality data and $i\lesssim80^{\circ}$, rotation curves are derived by fitting a tilted-ring model to the velocity field \citep{Begeman1989}; (ii) for galaxies with high-quality data and $i\gtrsim80^{\circ}$, rotation curves are derived using the envelope-tracing method \citep{Sancisi1979} because projection effects prohibit the use of a velocity field; (iii) for galaxies with data of lower quality, rotation curves are derived using major-axis position-velocity diagrams \citep{deBlok1996, Verheijen2001}, which are conceptually similar to long-slit spectroscopic observations. 

For many low-mass and LSB galaxies, we have combined H{\small I}/H$\alpha$ rotation curves: their inner rising portions are traced at higher spatial resolutions using long-slit and integral-field H$\alpha$ spectroscopy \citep{deBlok2001, KdN2006, KdN2008}. For the other galaxies, the inner rotation curves were derived taking resolution effects into account by using position-velocity diagrams \citep{deBlok1996, Verheijen2001} or building H{\small I} model-cubes \citep{Swaters2009, Lelli2012}. The data collection is described in more detail in the SPARC main paper \citep{SPARC}.

For all of these galaxies, we searched the digital archive of the {\it Spitzer Space Telescope} for images at 3.6 $\mu$m, which provide the closest link between observed luminosity and stellar mass. Several studies, indeed, suggest that the stellar mass-to-light ratio (M$_{*}$/L) is nearly constant at 3.6 $\mu$m over a broad range of galaxy masses and morphologies \citep{Martinsson2013, Meidt2014, McGaugh2014, Schombert2014b}. We found 176 objects with useful images and derived surface-brightness profiles following standard procedures \citep{Schombert2014a}. These surface-brightness profiles were then used to calculate the stellar contribution to the observed rotation curve, solving the Poisson equation for a disc with finite thickness \citep{Casertano1983}. Similarly, the gas contribution was calculated using radial gas density profiles from H{\small I} observations (corrected for the contribution of Helium). Once the contributions of gas and stars are determined, the DM contribution can be fitted to derive the parameters of the DM halo.

Our sample of galaxies is constructed to represent the broadest possible range of rotating disc galaxies. It contains galaxies spanning the range $3 \times 10^7 \lesssim M_* \,(\text{M}_{\odot}) \lesssim 3 \times 10^{11}$, $20 \lesssim V_{\text{rot}} \,(\text{km/s}) \lesssim 300$, and $3 \lesssim \Sigma_* \,(\text{M}_{\odot}/\text{pc}^{2}) \lesssim 1500$. Gas fractions from near zero to near unity are represented. While this sample is not statistically complete, it provides a much broader range of galaxy properties compared to other available samples.  In particular it has more low-mass and LSB galaxies than typical for rotation curve samples. This provides the widest view of galaxy properties currently available.

\section{Fitting the Rotation Curves}
\label{RCfit}
\subsection{Data Selection}
Rotation curves provide a faithful tracer of the gravitational potential, provided that the data are accurate and the galaxy is in equilibrium. To assure that the galaxies we analyse satisfy this criterion, we impose selection requirements. We require galaxies to be symmetric, sampled adequately, and not be too face-on.  

First, we require all galaxies to have at least five data points along the rotation curve which removes four galaxies.  The observed velocities have been corrected by $1/\sin(i)$ to account for the inclination of the disc on the sky plane. This is standard procedure, but becomes uncertain for face-on galaxies because $1/\sin(i)$ becomes large. We therefore exclude galaxies with $i < 30^{\circ}$. Most kinematic studies, from which this data set is built, apply this criterion during the sample selection or the data analysis; hence, only 16 galaxies are removed by this requirement. Clearly, this does not introduce any selection bias as galaxy discs are randomly oriented on the sky.  We also exclude nine galaxies with major kinematic asymmetries, since their rotation velocities are likely affected by strong non-circular motions and do not provide a faithful tracer of the gravitational potential. These restrictions reduce our initial sample from 176 galaxies to 147. It is unclear how galaxies with major asymmetries bias the sample, however these represent 5\% of the original data set and are unlikely to skew our results significantly. Minor asymmetries, indicative of small non-circular motions, are taken into account by the error bars on the rotation velocities, which are derived considering the differences between the approaching and receding sides of the disc \citep{Noordermeer2007, Swaters2009}. 

\subsection{Halo Models}
For each of the 147 different rotation curves, we fit both the DC14 and NFW models using an MCMC. For both models, we define the virial radius of the halo, R$_{vir}$, as the radius where the average halo density equals $\Delta$ times the critical density of the Universe where $\Delta=93.6$.  For this work, we choose $H_0=73$~km~s$^{-1}$ Mpc$^{-1}$. The halo virial velocity is then given by 
\begin{equation}
V_{vir} = \sqrt{G M_{vir}/R_{vir}}.
\end{equation}
We also define the halo concentration as 
\begin{equation}
c_{vir}=\text{R}_{vir}/r_{-2}, 
\end{equation}
where $r_{-2}$ is the radius at which the logarithmic slope of the density profile is $-2$. Since the NFW and DC14 profile have different inner slopes, the NFW concentration, $c_{vir,\rm NFW}$, is a different quantity than the DC14 concentration, $c_{vir,\rm DC14}$. We can convert between $c_{vir,\text{DC14}}$ and $c_{vir,\text{NFW}}$ using the following relation \citep{DC2014}:
\begin{equation}
c_{vir,\text{NFW}} = c_{vir,\text{DC14}}/(1.0+e^{0.00001[3.4(X+4.5)]})
\end{equation}
which gives a primordial NFW halo for every galaxy modelled with the DC14 profile\footnote{The coefficient in the exponent we use here (0.00001) is different from that given in \cite{DC2014} (0.00003), as this new value was found to provide a better fit to the simulation data.}.

The NFW density profile is given by
\begin{equation}
\rho_{\text{NFW}}(r)=\frac{\rho_{\text{s}}}{\left(\frac{r}{r_{\text{s}}}\right)\left[1+\left(\frac{r}{r_{\text{s}}}\right)\right]^2}
\end{equation}
where $\rho_{\rm s}$ and $r_{\rm s}$ are scale density and scale radius, respectively.
Next, 
\begin{equation}
\rho_{\text{s}}=\frac{\text{M}_{vir}}{4\pi r_s^3[\ln(1+c_{vir})-\frac{c_{vir}}{1+c_{vir}}]}
\end{equation}

The DC14 density profile is given by
\begin{equation}
\rho_{\text{DC14}}(r)=\frac{\rho_{\rm s}}{\left(\frac{r}{r_{\rm s}}\right)^{\gamma}\left[1+\left(\frac{r}{r_{\text{s}}}\right)^{\alpha}\right]^{(\beta-\gamma)/\alpha}}.
\end{equation}
where $-\gamma$ and $-\beta$ give the inner and outer logarithmic slopes, respectively, and $\alpha$ describes the transition between the two. These are not free parameters, but they are related to the M$_{*}$/M$_{\rm halo}$ ratio by
\begin{equation}
\label{DC14eqn}
\begin{aligned}
&\alpha=2.94-\log_{10}[(10^{X+2.33})^{-1.08}+(10^{X+2.33})^{2.29}]\\
&\beta=4.23+1.34X+0.26X^2\\
&\gamma=-0.06-\log_{10}[(10^{X+2.56})^{-0.68}+(10^{X+2.56})].
\end{aligned}
\end{equation}
where $X=\log_{10}(\text{M}_{*}/\text{M}_{\text{halo}})$. For the DC14 model, one has
\begin{equation}
r_{-2}=\left(\frac{2-\gamma}{\beta-2}\right)^{1/\alpha}r_{\text{s}}.
\end{equation}
In this formalism, the NFW profile has ($\alpha$, $\beta$, $\gamma$) = (1, 3, 1) and $r_{-2} = r_{\rm s}$.  The DC14 model predicts that for $-3.75\lesssim\text{M}_*/\text{M}_{\text{halo}}\lesssim-1.75$, expansion is the dominant process that regulates the transformation of primordial DM haloes.  The simulations used to derive the DC14 model neglect the effects of supermassive black holes, which may affect both inflows and outflows of gas, and therefore the predictions are likely accurate up to galaxies with M$_{\text{halo}}\sim10^{12}$~M$_{\odot}$.  Furthermore, these simulations only sample galaxies with $\log_{10}(\text{M}_*/\text{M}_{\text{halo}})<-1.3$.  For this reason, we use the limiting values of $\alpha$, $\beta,$ and $\gamma$ at $X=-1.3$ for $X>-1.3$ and simply denote that the model has been extrapolated for galaxies with total mass greater than $10^{12}$ M$_{\odot}$.

Using these density profiles, we calculated the DM contribution to the total rotation curve, V$_{\text{DM}}(r)$, assuming spherical symmetry.  In order to find the total resulting rotation curve and to compare with observations, we add V$_{\text{DM}}(r)$ to the stellar and gas components as follows:
\begin{equation}
\text{V}_{\text{c}}(r)=\sqrt{\text{V}_{\text{DM}}(r)^2 + \text{V}_{\text{gas}}(r)^2 +(\text{M$_{*}$/L})\text{V}_{\text{stars}}(r)^2}
\end{equation}

\subsection{MCMC Fitting}
\label{MCMCfit}
We use the parallel-tempered ensemble sampler from the open source Python package, {\small{emcee}} \citep{emcee}, to map the posterior distributions of three free parameters: $\log_{10}(\rm V_{vir})$, $\log_{10}(c_{\rm vir})$, and $\log_{10}({\rm M_*/L})$.  This sampler better deals with multimodal posteriors compared to the standard ensemble sampler.  By sampling in comparable parameters, the MCMC sampler estimates the posterior probability distribution equivalently.  For all models, we define a fiducial set of priors.  We impose extremely loose priors on the first two parameters, $10.0 <  \text{V}_{vir} < 500.0$~km/s and $1.0<c_{vir}<100.0$ (these are imposed as flat priors in log-space) to ensure fast computational speed.  These loose priors have no physical basis but are well beyond the regime of parameters that real galaxies in our sample are expected to exhibit and only effect the NFW model fits.  We place a constraint on M$_{*}$/L such that $0.3< \text{M$_{*}$/L} < 0.8$, as suggested by stellar population synthesis models \citep{Meidt2014, McGaugh2014, Schombert2014b}. We impose a further condition for all models, $(\text{M}_*+\text{M}_{\text{gas}})/\text{M}_{\text{halo}}<0.2$, so that the baryon fraction is always less than the cosmological value.   

The MCMC chains are initialized with 100 walkers for each of the 20 different temperatures and their starting positions are randomly assigned within the logarithmically uniform flat priors.  The chains are then run for a burn in period of 500 iterations.  The sampler is then reset and run for an additional 1,000 steps.  We have checked that the acceptance fraction is between 10\% and 50\% for all galaxies.  In order to obtain acceptance fractions in this range, we have tuned the {\small{emcee}} $a$ parameter, which controls the size of the stretch-move, and in general $a=3$ is sufficient for our purposes.  We mandate that the chains are run for a minimum of 10 autocorrelation times although the default 1,000 iterations for each galaxy are often much more than sufficient.  We use the Gelman-Rubin statistic (R) and require that $|R-1|<0.1$, assuming the walkers are independent samples, for each free variable chain to ensure convergence for our lowest temperature walkers.  We find that out of the 588 independent simulations, 586/588 have $|R-1|<0.01$.  The two that do not satisfy this stricter criteria have very small secondary modes that do not affect the halo parameter estimation.  The final 3D posterior distribution is constructed from the 1,000 iterations of each walker at the lowest temperature for a total of 100,000 unique samples and we record the maximum likelihood for each model.  3D error bars are calculated using the open source {\small{GetDist}}\footnote{We have used the stand-alone version available at https://github.com/cmbant/getdist that was originally packaged with {\small{CosmoMC}} \citep{Lewis2002}} software based on the 95\% confidence interval from the ``margestats" output. 

Some of the posteriors obtained are multimodal.  This is much more common for DC14 than for NFW.  For the DC14 model, there are two values of $X$ in Equation~\ref{DC14eqn} that can give the same inner slope and some of the rotation curves in our sample do not probe out to ${\rm V_{flat}}$ leading to a degeneracy.  Furthermore, certain features that may appear in the rotation curves may lead to degeneracies in parameter space for either the DC14 or NFW models.  We separate the modes by using the HDBSCAN\footnote{https://github.com/lmcinnes/hdbscan} algorithm \citep{Campello2013} to find clusters of points in the posteriors.  We run this on each of the galaxy posteriors.  If only one cluster is found, all 100,000 points are returned as a single mode which is then passed to {\small{GetDist}}.  Alternatively, if more than one mode is found, the 100,000 points are split into their corresponding modes and passed to {\small{GetDist}} separately.  We only consider modes that represent $>5\%$ of the posterior (i.e. sampled by more than 5,000 points).

\section{Results}
\label{results}
\subsection{Goodness of Fit to the Rotation Curves}
Using the loose priors described in Section~\ref{MCMCfit}, we have fit the 147 rotation curves in our sample with both the DC14 and NFW halo models. To highlight the differences between the two models, Figure~\ref{RCcomp} shows two example galaxies that differ in both stellar mass and surface-brightness. For the high-mass, high-surface-brightness galaxy (NGC 5055, $L_{[3.6]} = 10^{11}$ $L_{\odot}$, $\Sigma_{\rm eff} = 1100$ $L_{\odot}$ pc$^{-2}$), the DC14 and NFW models are nearly indistinguishable because baryonic processes are not able to erase the primordial cusp. For the low-mass, LSB galaxy (NGC 3109, $L_{[3.6]} = 10^8$ $L_{\odot}$, $\Sigma_{\rm eff} = 6$ $L_{\odot}$ pc$^{-2}$), the inner structure of the halo density profile is much better reproduced by DC14 than NFW because baryonic processes have transformed the inner cusp into a core. These results are representative for the general trends between high and low-mass galaxies in SPARC.

\begin{figure}
\centerline{\includegraphics[scale=0.5]{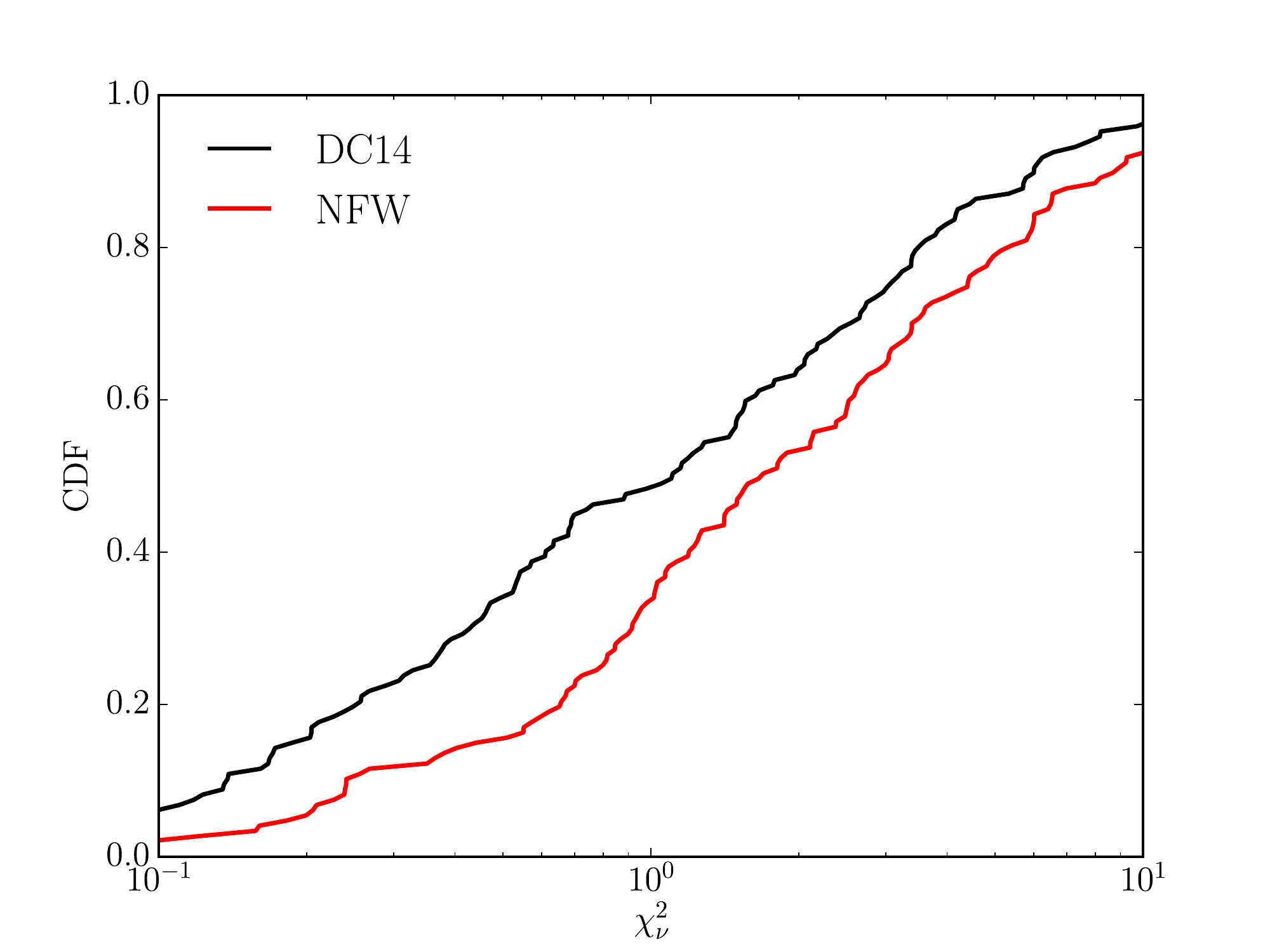}}
\caption{Cumulative distribution function (CDF) of the $\chi_{\nu}^2$ values of the maximum posterior rotation curve fits for the DC14 model compared with NFW.}
\label{x2cdf}
\end{figure}

In Figure~\ref{x2cdf}, we plot the cumulative distribution function (CDF) of the reduced $\chi^2$ ($\chi_{\nu}^2$) values of the maximum posterior fits to the rotation curves. The vast majority of galaxies fitted with the DC14 model have $\chi_{\nu}^2 < 1.5$, while this is not the case for NFW. Over the whole sample the $\chi_{\nu}^2$ is significantly smaller for the DC14 model than for NFW. The median $\chi_{\nu}^2$ is 1.11 for DC14 while $\chi_{\nu}^2 = 1.69$ for NFW. This sample of galaxies is clearly better fit by haloes with modified inner structure compared to the NFW profile.

The median $\chi_{\nu}^2$ for the DC14 model is very close to unity, suggesting that this model is not only superior to NFW in fitting the data but also may be closer to capturing the true DM density profiles of real galaxies. Several galaxies fit by the DC14 model, however, have a maximum posterior $\chi_{\nu}^{2}< 1$ as well as $\chi_{\nu}^2> 1$. Naively, one may conclude that the error bars on some individual rotation curves are either over- or under-estimated. We should keep in mind, however, that $\chi_{\nu}^2$ is a probability distribution; hence, we do expect that several fits across the whole sample will fall into these two regimes. The shape of the $\chi_{\nu}^2$ CDF of the whole sample for both models still differs from the shape of the expected distribution. There are several possible explanations for this since the world of astronomical data is far from perfect. The error bars on the rotation curves do not include systematic uncertainties due to the assumed distance, which changes from galaxy to galaxy. The vertical density distribution of stellar disks may not be described by a single exponential (as assumed here) and DM haloes may not be spherical. These systematic uncertainties are expected to be small, but likely play a role in the $\chi_{\nu}^2$ distribution.  

Nevertheless, we stress that both halo models were fit to the same galaxies suffering from the same systematics. Therefore one can address how different halo models compare to one another on a galaxy by galaxy basis. The results are solid and unmistakable: the DC14 model fits the data better than NFW in nearly all cases. Given that our sample is respectably large and covers a very broad range in galaxy properties, it is fair to conclude that the DC14 halo model represents the properties of real galaxies better than NFW. This is further investigated in Section~\ref{MCLCDMP}.

\begin{figure*}
\centerline{\includegraphics[scale=0.5]{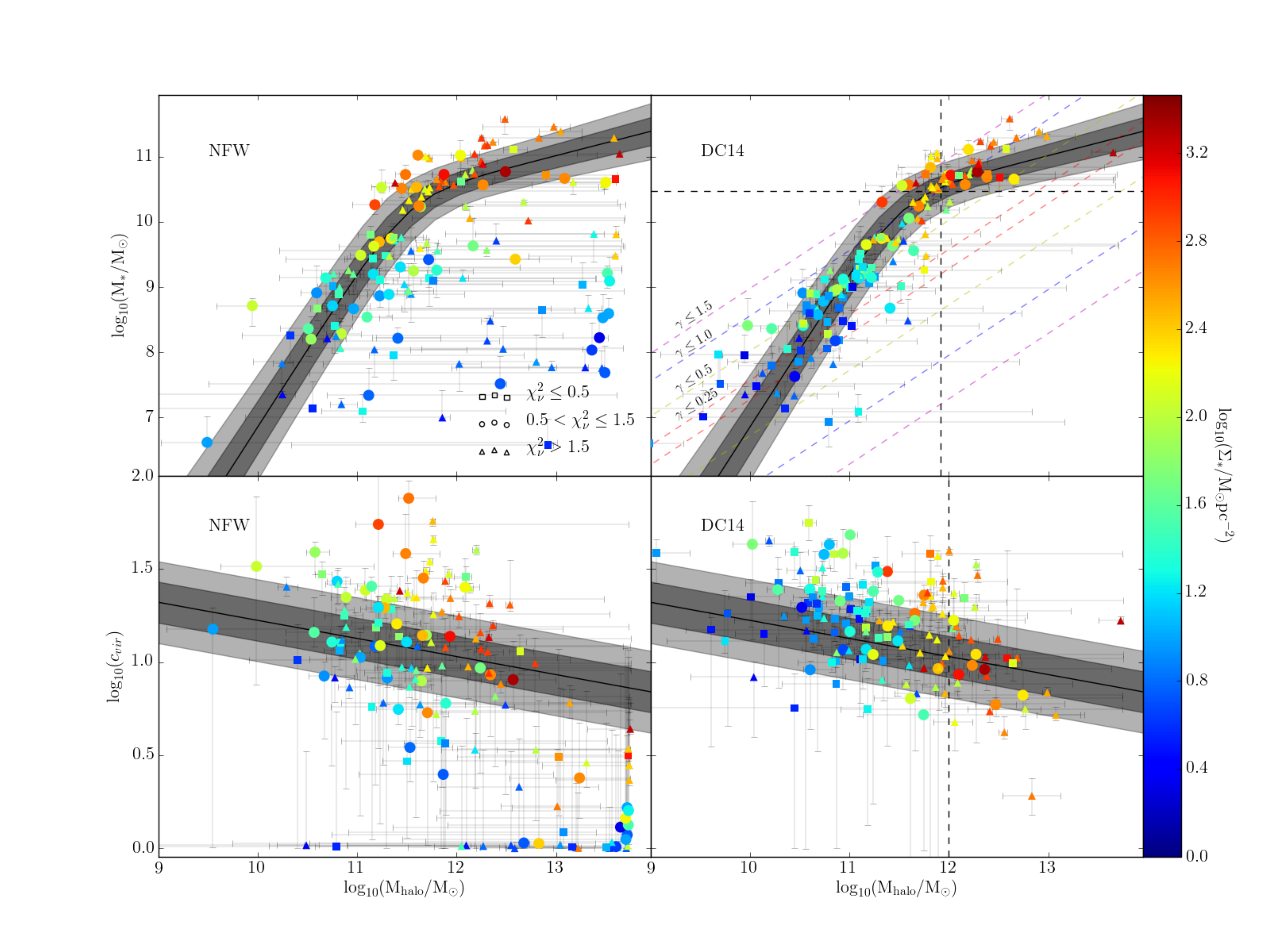}}
\caption{Maximum posterior NFW (left) and DC14 (right) halo fits compared to the abundance matching \protect{\citep{Moster2013}}(top) and mass-concentration relations\protect{\citep{Dutton2014}} (bottom).  Note that halo mass in the top row represents M$_{200}$ while halo mass in the bottom row represents M$_{vir}$.  The black lines represent the mean relation while the dark and light grey shaded regions show the $1\sigma$ and $2\sigma$ scatter, respetively.  The points are coloured by $\log_{10}(\Sigma_*/\text{M}_{\odot}\text{pc}^{-2})$ where $\Sigma_*$ is the central surface-brightness of the stars in the galaxy.  The coloured dashed lines in the top-right panel depict lines of constant inner slope for the DC14 model for galaxies of average concentration.  The black vertical dashed lines in the right-hand panels show where the DC14 model is extrapolated outside the range of halo and stellar masses used to predict it.  Error bars represent the projected 95\% confidence interval of the posterior probability distribution.  Only the cosmologically motivated parameters are shown for multimodal posteriors.}
\label{quadflat}
\end{figure*}

\subsection{Comparison with $\Lambda$CDM expectations}
With such a large sample of galaxies, we can test these models in a cosmological context.  The $\Lambda$CDM cosmogony predicts distinct relations for populations of DM haloes.  Specifically, the fitted DM haloes must be consistent with both the M$_*$-M$_{\text{halo}}$ relation from abundance matching \citep{Moster2013} and the mass-concentration relation from cosmological, DM-only simulations \citep{Dutton2014}.

The top panel of Figure~\ref{quadflat} shows the M$_*$-M$_{\text{halo}}$ relation obtained from the maximum posterior rotation curve fits that provide the optimal M$_{\text{halo}}$ for both halo models\footnote{Note that when comparing with both the abundance matching and mass-concentration relations, we account for the fact that DM-only simulations were used to calibrate these relations by scaling the relations by the appropriate baryon fraction.  Furthermore, we scale the halo masses to the appropriate $\Delta$ for the given cosmology of each of these relations (M$_{200}$ corresponds to where the average density inside the halo is $200\rho_{\rm crit}$ whereas M$_{vir}$ corresponds to where the average density inside the halo is $\Delta\rho_{\rm crit}$ and $\Delta$ is taken from the cosmology used in deriving the relevant relation).}.  The best-fitting parameters of the NFW haloes do not adhere to the M$_*$-M$_{\text{halo}}$ relation required by abundance matching.  Many galaxies scatter below the 2$\sigma$ contours of the predicted relation, or are pinned to the edge of the priors.  This indicates that the NFW density profile fails in two distinct aspects: (i) it does not provide good fits to rotation curves when compared to DC14, and (ii) the fitted haloes are inconsistent with fundamental $\Lambda$CDM expectations.  Conversely, galaxies fitted with the DC14 model follow the expected relation from abundance matching reasonably well within the error bars, with only $\sim10\%$ of outliers falling off the M$_*$-M$_{\text{halo}}$ relation.  

A known shortcoming of NFW halo fits is that the concentration is often too low, especially for galaxies of LSB \citep{deBlok2001,KdN2009}. This can be seen in the bottom left panel of Figure~\ref{quadflat} as the cluster of points much lower than the mass-concentration relation.  If we insist that NFW must be the correct halo form, we then fail to find that most of the halo parameters fall within the $2\sigma$ contours of the mass-concentration relation predicted by $\Lambda$CDM.  For the DC14 halo, we can relate the halo parameters back to their original form, before baryonic infall, and compare them with the mass-concentration relation from DM-only simulations.  In the bottom-right panel of Figure~\ref{quadflat}, we see that significantly more points for the DC14 model fall on this relation compared to the NFW model.  The feedback-softened density profiles of DC14 haloes allow for fits with concentrations that are broadly consistent with the expected relation within the error bars.

We note that the observed rotation curves typically extend out to $\sim10\%$ of the virial radius, limiting our ability to constrain the total mass or the virial velocity of the halo.  This can lead to large degeneracies between parameters such as mass and concentration as halo models may look similar in the inner regions of the galaxy but differ at larger radii.  This can be seen as large error bars on the points in Figure~\ref{quadflat}.  Note however that the fitting parameters are not orthogonal and thus the region mapped out by the error bars is over representative of the true posterior mapped by these parameters which is often in the shape of a much thinner crescent.  In Appendix~\ref{rcplots}, we describe the full posteriors and rotation curve fits for a small subsample of galaxies, which represents the diversity among the SPARC data set.

\begin{figure*}
\centerline{\includegraphics[scale=0.5]{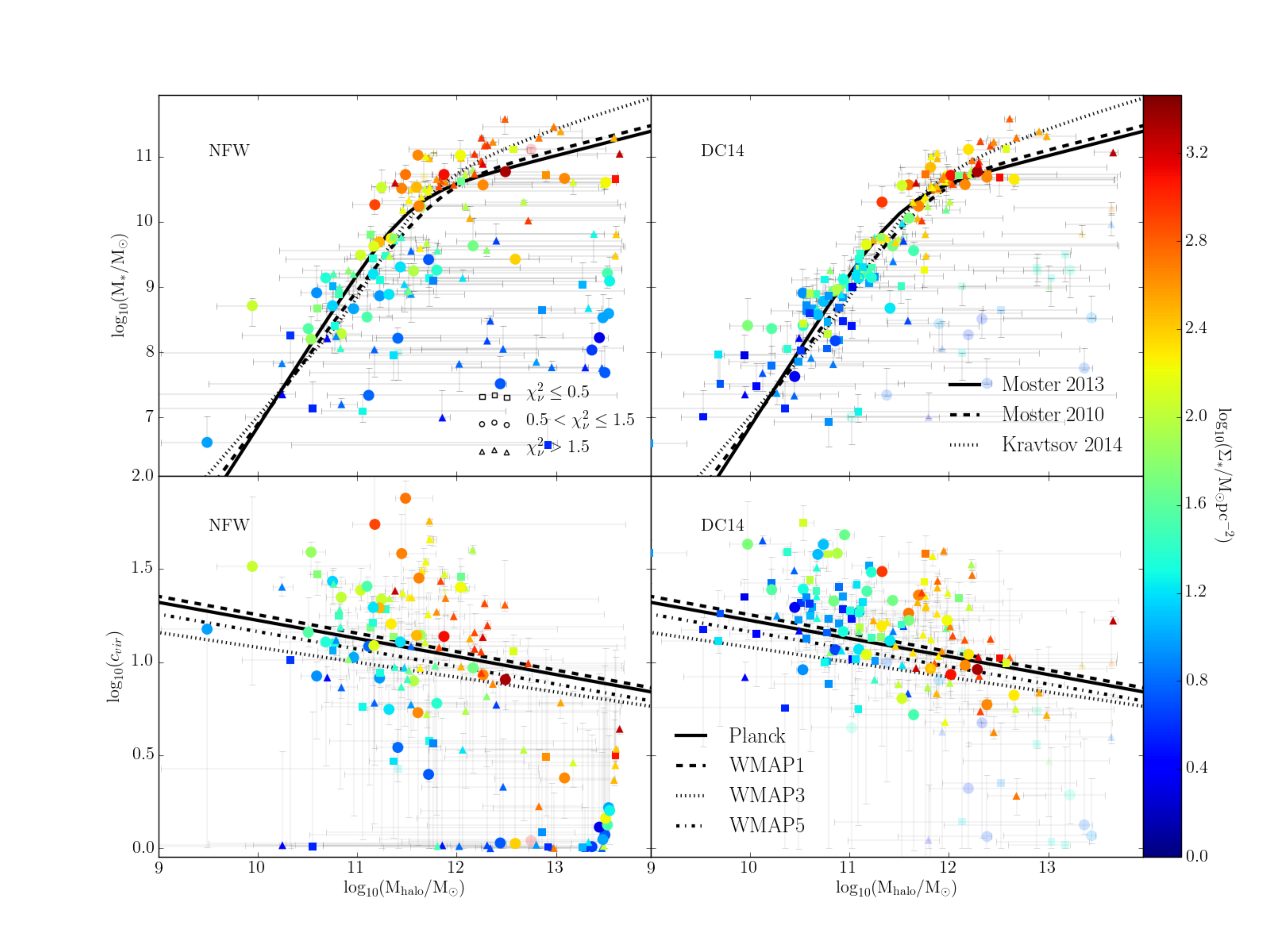}}
\caption{Maximum posterior NFW (left) and DC14 (right) halo fits compared to various abundance matching (top) and mass-concentration relations (bottom).  Note that halo mass in the top row represents M$_{200}$ while halo mass in the bottom row represents M$_{vir}$.  The points are coloured by $\log_{10}(\Sigma_*/\text{M}_{\odot}\text{pc}^{-2})$ where $\Sigma_*$ is the central surface-brightness of the stars in the galaxy.  Error bars represent the projected 95\% confidence interval of the posterior probability distribution.  The lighter, more translucent points in all panels represent the secondary modes for galaxies that have a multimodal posterior}
\label{multiflat}
\end{figure*}

In Figure~\ref{multiflat}, we show the maximum posterior fits for both the NFW and DC14 models against other measurements of the M$_*$-M$_{\text{halo}}$ and mass-concentration relations.  In particular, we consider the M$_*$-M$_{\text{halo}}$ relations from \cite{Moster2010} and \cite{Kravtsov2014} and the mass-concentration relations for WMAP1, WMAP3, and WMAP5 cosmologies from \cite{Maccio2008}.  At lower masses, all three M$_*$-M$_{\text{halo}}$ relations agree reasonably well and thus we come to a similar conclusion: the DC14 model tends to follow the relations better than NFW.  At the high-mass end, the \cite{Kravtsov2014}  M$_*$-M$_{\text{halo}}$ relation deviates from the others and has a steeper slope that seems to agree slightly better with the maximum posterior halo parameters.  This, however, is in the extrapolated regime for the DC14 halo model and we refrain from making any significant conclusions based on this data alone.  The mass-concentration relations from WMAP1 and WMAP5 tend to agree reasonably well within $\sim20\%$ of latest results from Planck.  The WMAP3 relation falls slightly below the others and we find that our DC14 halo model points tend to scatter slightly higher than this relation.  This cosmology has a low value of $\Omega_m$ and $\sigma_8$ which means that haloes assemble later and have lower concentrations \citep{Maccio2008}.  The DC14 model is, in principle, independent of cosmology so it is not clear which cosmology is the best for comparison.  We use Planck as our default relation for comparison as this represents the most recent and accurate investigation.  Nevertheless, when these relations are imposed as priors, we see that the difference in $\chi^2_{\nu}$ is not significant between WMAP3 and Planck cosmologies.

\subsection{Goodness of fit to the rotation curves with $\Lambda$CDM priors}
\label{MCLCDMP}
In order to better constrain model parameters, we can test how the fits to the observed rotation curves change when the M$_*$-M$_{\text{halo}}$ and mass-concentration relations are imposed as priors in our MCMC analysis.  These relations are imposed as lognormal priors with means and uncertainties as given by \cite{Moster2013} and \cite{Dutton2014} and are combined with the fiducial priors listed in Section~\ref{MCMCfit}.  The MCMC chains are then rerun for each of the 147 individual galaxies for both the DC14 and NFW haloes.  

\begin{figure*}
\centerline{\includegraphics[scale=0.5]{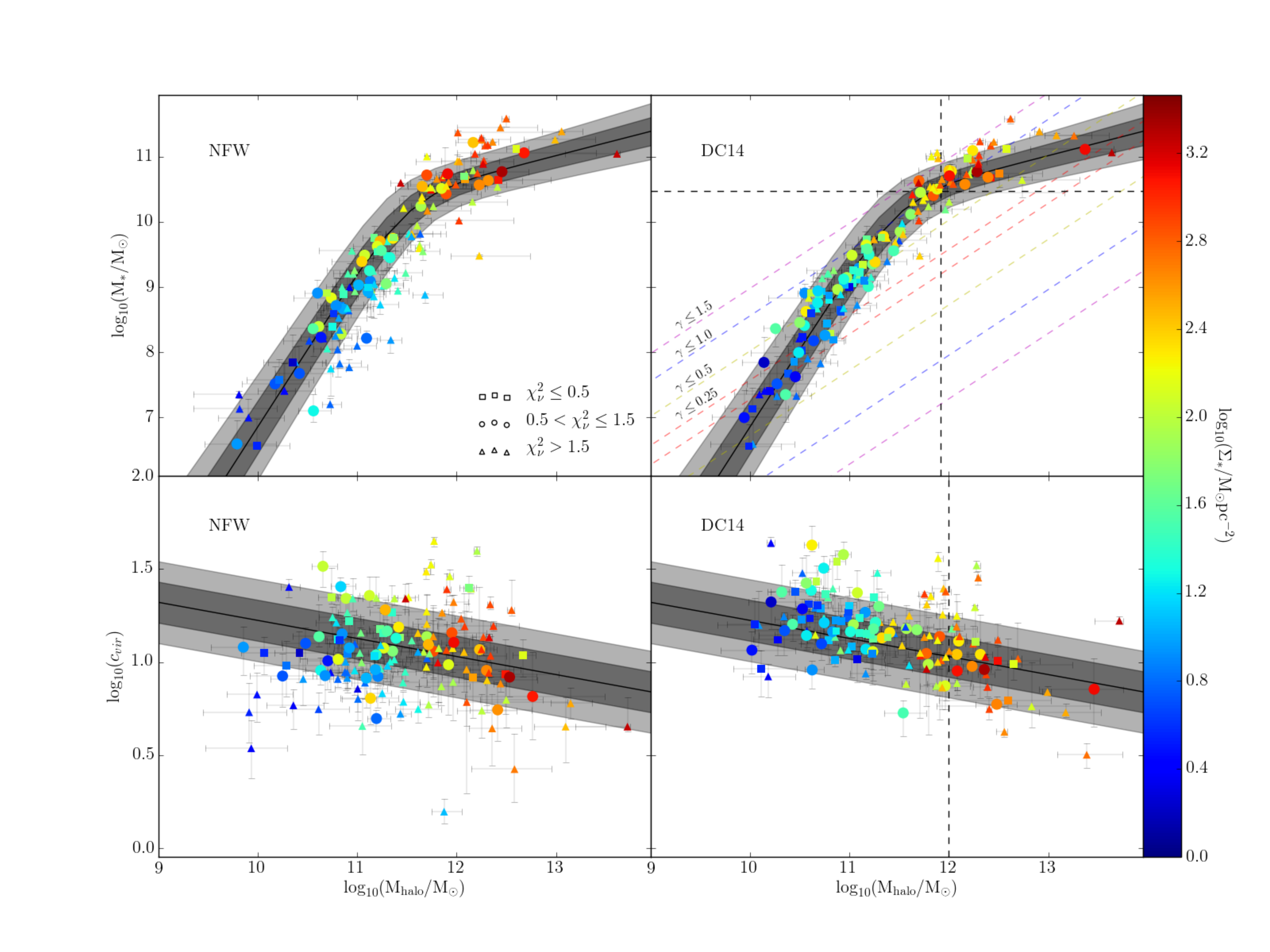}}
\caption{Maximum posterior NFW (left) and DC14 (right) halo fits compared to the abundance matching \protect{\citep{Moster2013}}(top) and mass-concentration relations\protect{\citep{Dutton2014}} (bottom) when the $\Lambda$CDM priors are imposed.  Note that halo mass in the top row represents M$_{200}$ while halo mass in the bottom row represents M$_{vir}$.  The black lines represent the mean relation while the dark and light grey shaded regions show the $1\sigma\sigma$ and $2\sigma$ scatter, respetively.  The points are coloured by $\log_{10}(\Sigma_*/\text{M}_{\odot}\text{pc}^{-2})$, where $\Sigma_*$ is the central surface-brightness of the stars in the galaxy.  The coloured dashed lines in the top-right panel depict lines of constant inner slope for the DC14 model for galaxies of average concentration.  The black vertical dashed lines in the right-hand panels show where the DC14 model is extrapolated outside the range of halo and stellar masses used to predict it.  Error bars represent the projected 95\% confidence interval of the posterior probability distribution.  Only the cosmologically motivated parameters are shown for multimodal posteriors.}
\label{quadlcdm}
\end{figure*}

In Figure~\ref{quadlcdm}, we show the maximum posterior fits compared to the abundance matching and mass-concentration relations predicted by $\Lambda$CDM.  Unsurprisingly, with these relations imposed, the majority of galaxies fitted with the DC14 halo profile adhere very well to the 2$\sigma$ contours of the predicted relations (see Appendix \ref{app1} for how these relations compare with inner density slope of the DC14 model).  There clearly still remains significantly more scatter around the relations when fitting the galaxies with the NFW model.  

Of course, with a strong enough prior, any relation can be recovered and what we aim to address here is how the quality of the rotation curve fits change with these relations imposed.  In this case, the median $\chi^2_{\nu}$ of the maximum posterior rotation curve fits over the entire sample is 1.33 for DC14 and 2.37 for NFW, indicating that DC14 provides much better fits to the rotation curves compared to NFW.  Remarkably, the median $\chi^2_{\nu}$ changes much less for the DC14 model when the $\Lambda$CDM priors are imposed compared to NFW, suggesting that the overall best-fitting halo parameters for the rotation curves are much more consistent with $\Lambda$CDM for the DC14 model, than for NFW.  This empirical observation arises because significantly more of the DC14 halo fits fall within the $2\sigma$ contours of the $\Lambda$CDM relations compared to NFW before the $\Lambda$CDM priors are included in the MCMC analysis.  With the relations imposed, 53\% of galaxies in our sample have $\chi^2_{\nu}<1.5$ for DC14, while only 37\% of galaxies have $\chi^2_{\nu}$ this low when fitted with the NFW model.  

\begin{figure*}
\centerline{\includegraphics[scale=0.5]{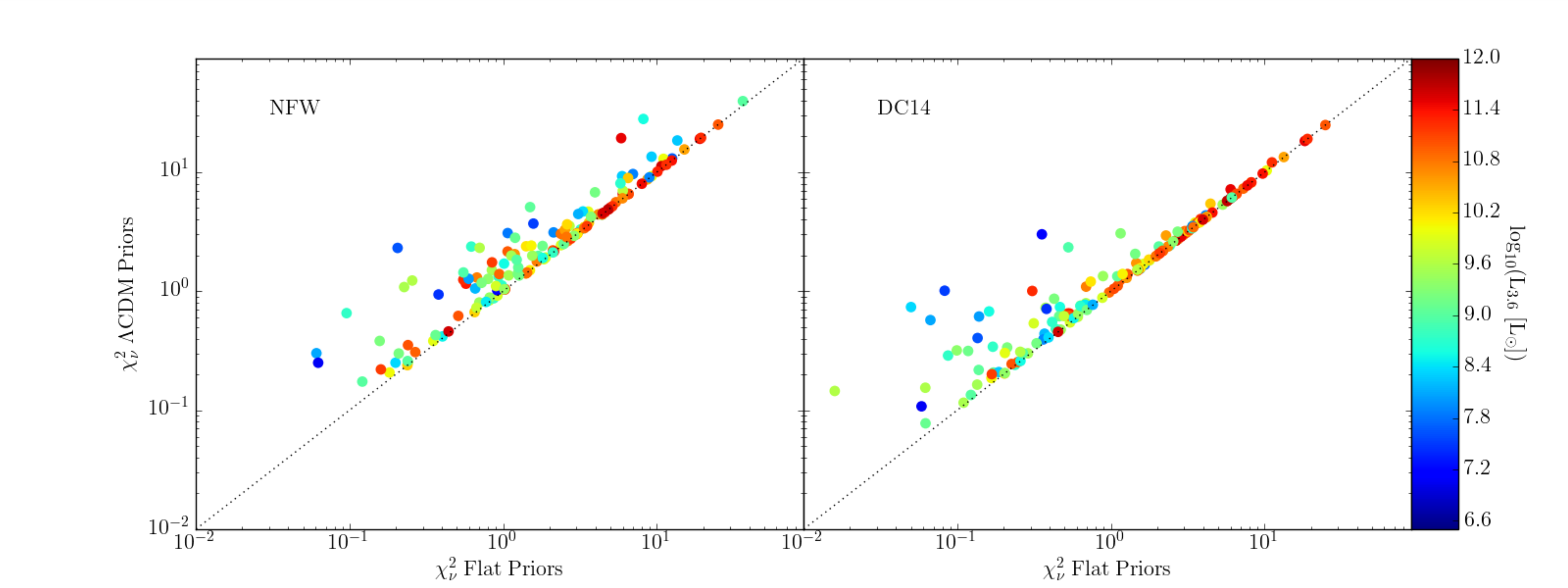}}
\caption{Maximum posterior $\chi^2_{\nu}$ of the rotation curve fitted with the fiducial set of priors compared to the maximum posterior $\chi^2_{\nu}$ of the rotation curve fitted with the $\Lambda$CDM priors for the NFW (left) and DC14 (right) models.  The points are coloured by their 3.6$\mu$ luminosity.  The dashed lines represent the one-to-one line of $\chi^2_{\nu}$, which is the minimum possible value of $\chi^2_{\nu}$, the galaxy fit can have when the $\Lambda$CDM priors are imposed.}
\label{x2rat}
\end{figure*}

In Figure~\ref{x2rat}, we plot the $\chi^2_{\nu}$ of the rotation curve fits with the fiducial priors versus the $\chi^2_{\nu}$ of the rotation curve fits with the $\Lambda$CDM priors.  Points which fall close to the diagonal dotted line have a $\chi^2_{\nu}$ which remains the same when the $\Lambda$CDM priors are imposed.  This means that, for these galaxies, the maximum posterior halo parameters derived from the fits with the fiducial set of priors are consistent with the expectations from $\Lambda$CDM.  From Figure~\ref{x2rat}, it is clear that many more points fall on the dotted line for the DC14 model compared to NFW as expected from Figure~\ref{quadflat}.  For the DC14 model, most of the galaxies that have a significant difference in $\chi^2_{\nu}$ between the fits with and without $\Lambda$CDM priors started with a $\chi^2_{\nu}<1$.  The values then increase when the $\Lambda$CDM priors are imposed but the new $\chi^2_{\nu}$ values still tend to stay below 1.  This means that they are still well fit and the median $\chi^2_{\nu}$ across the entire sample does not change.  This is not true for NFW, as points fall above the dotted line at all values of $\chi^2_{\nu}$ on the $x$-axis.  For the NFW halo model, the increase in $\chi^2_{\nu}$ tends to place the new value above the median which is why we see a significant increase in the median $\chi^2_{\nu}$ with the $\Lambda$CDM priors imposed.

We have coloured the points in Figure~\ref{x2rat} by their 3.6$\mu$ luminosity and we can see that there is a slight tendency for the more luminous galaxies to fall at higher values of $\chi^2_{\nu}$.  This is due to the fact that the more luminous galaxies are in general larger and tend to have more rotation curve points compared to the lower luminosity galaxies.  Some of these galaxies have features in their rotation curves that will not be well captured by any smooth halo model and thus the galaxies tend to have slightly higher values of $\chi^2_{\nu}$.  This, however, affects both the DC14 model and NFW equally and for many of these galaxies, the DC14 model is in a parameter space where the density profile is similar to NFW.  Furthermore, we show in Appendix~\ref{rcplots} that the general trends in these high luminosity galaxies tend to be well captured by the rotation curve fits despite having slightly higher $\chi^2_{\nu}$ values.

We must be aware that the exact distribution of $\chi^2_{\nu}$ is dependent on the ratio of high-mass to low-mass galaxies in the sample.  Although our sample is representative and spans a wide range in stellar mass, surface-brightness, and circular velocity, it is not a volume limited sample.  To understand how each of these models would perform if fitted to a truly complete volume limited sample, we create 10,000 Monte Carlo resampled catalogues of galaxies where the probability of selecting each galaxy from the parent sample is based on the relative number density of galaxies of that magnitude given by the $3.6\mu$m luminosity function of late-type galaxies \citep{Dai2009}.  The average median values of $\chi^2_{\nu}$ over the 10,000 catalogues is $0.72\pm0.04$ for the DC14 model and $1.9\pm0.28$ for NFW.  This suggests that a true volume limited sample, which is dominated by low-mass galaxies, would be much better fitted by the DC14 model than NFW.

\subsection{Bayesian Model Comparison}
\begin{figure}
\centerline{\includegraphics[scale=0.5]{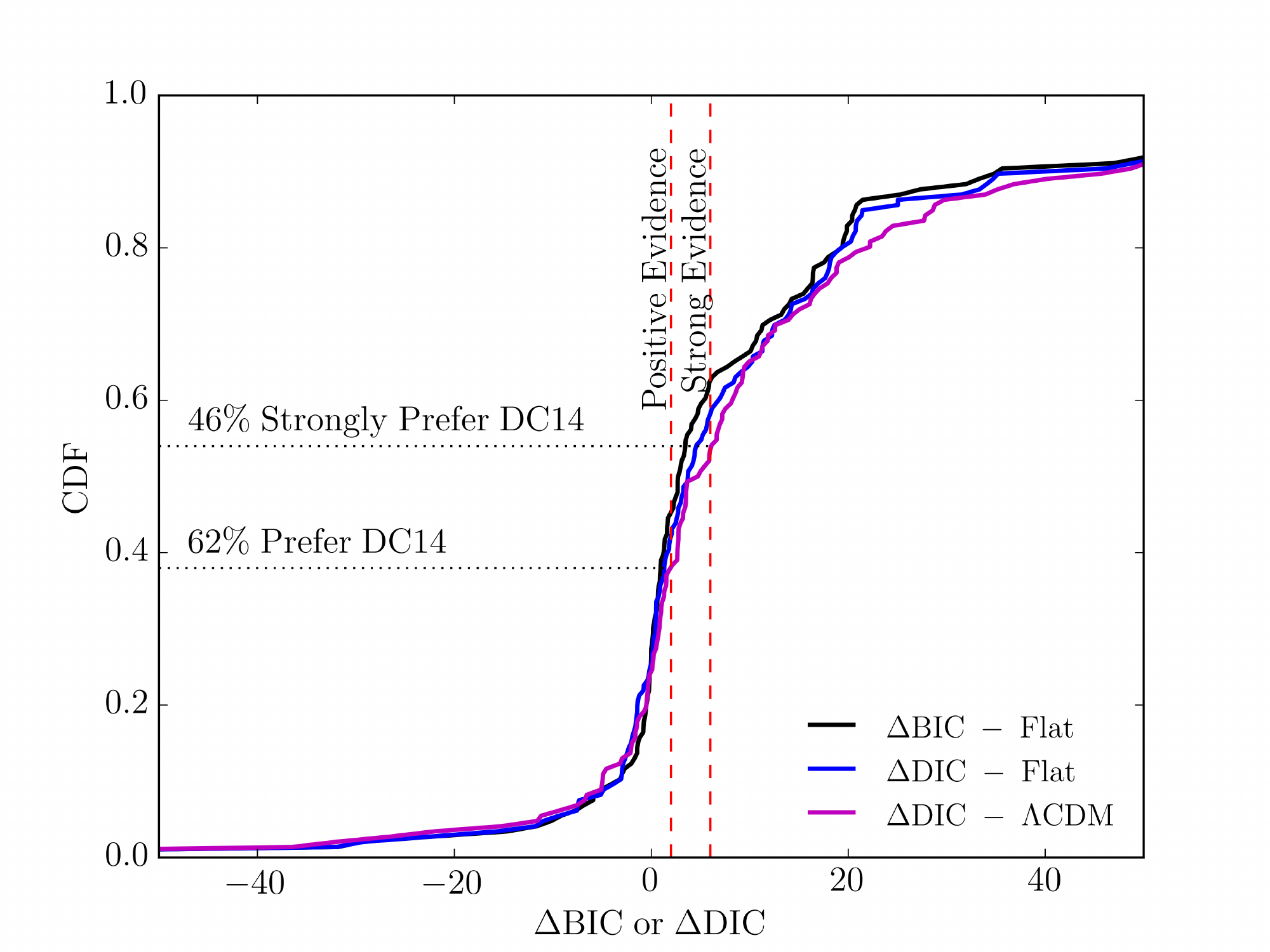}}
\caption{CDF of the $\Delta$BIC and $\Delta$DIC values for the halo fits for models with and without abundance matching and mass-concentration priors.  The two vertical lines represent the threshold beyond which the DC14 model is either favoured or strongly favoured.  The majority of our sample demonstrates evidence in support of the DC14 model over the NFW model regardless of whether $\Lambda$CDM priors are included.  The results are consistent between the $\Delta$BIC and $\Delta$DIC.  The percentages quoted on the plot show the values for the $\Delta$DIC for the model that includes the $\Lambda$CDM priors.}
\label{BIC}
\end{figure}

One can quantify the improvement of the DC14 model over NFW, by using the Bayesian information criteria,  
\begin{equation}
{\rm BIC}=-2\ln\mathcal{L}+k\log(n),
\end{equation}
where $k$ is the number of parameters, $n$ is the number of data points, and $\mathcal{L}$ is the likelihood of the maximum likelihood point, \citep{BIC} to select between the models.  In Figure~\ref{BIC}, we plot the CDF of the $\Delta$BIC for our entire sample for the fits without the $\Lambda$CDM priors.  We find that 54\% of our galaxies have a $\Delta$BIC~$>2$, which is positive evidence that DC14 is preferred over NFW, while 37\% have $\Delta$BIC~$>6$ indicating strong evidence for preference of DC14 \citep{kassr95}.  In the majority of cases with no model preference (i.e. $-2<\Delta$BIC~$<2$), the DC14 profile is observationally indistinguishable from the NFW profile because $\log_{10}(\text{M}_*/\text{M}_{\text{halo}}) > -1.75$, which corresponds to M$_*\geq1.8\times10^{10}$~M$_{\odot}$ for M$_{\text{halo}}=10^{12}$~M$_{\odot}$.  Finally, 13\% of galaxies in the sample have a $\Delta$BIC~$<-2$, which indicates a preference for the NFW model; however, in the majority of these cases (74\%), neither model provides a particularly good fit to the observed rotation curves in terms of the $\chi^2_{\nu}$.  Overall, this indicates a statistically significant preference for the feedback-softened DC14 model over NFW.

The $\Delta$BIC only considers the maximum likelihood point in our sample, and from a Bayesian perspective, one should use a statistic that considers the entire posterior.  For this we turn to the Deviance Information Criteria (DIC) defined as 
\begin{equation}
{\rm DIC}\equiv D(\bar{\theta}) + 2p_D.  
\end{equation}
Here, $p_D=\overline{D(\theta)}-D(\bar{\theta})$ and $D(\theta)=-2\ln \mathcal{L}+C$, where $C$ is a data-dependent constant that will vanish from a derived quantity \citep{Spiegelhalter2002,Liddle2007}.  The DIC can be easily calculated from a set of MCMC chains and we calculate the $\Delta$DIC between the NFW and DC14 models for the runs both with and without priors.  In the case of multimodal posteriors, we consider the cosmologically motivated mode.  

There is some debate in the literature on which value to take for $\bar{\theta}$ since it can affect the statistic \citep{Spiegelhalter2002,Celeux2006,Liddle2007}, and here we set $\bar{\theta}$ to the mean values of the 1D marginalized posteriors on each of our fitting parameters.  We have checked the results using the maximum posterior point as well and we find that the results are robust to within a few percent.  

For the runs with our fiducial set of flat priors, we find that $58\%$ of galaxies show evidence in favour of the DC14 model ($\Delta$DIC $>2$) while $41\%$ of galaxies show strong evidence for preference of the DC14 model ($\Delta$DIC $>6$).  On the contrary, $16\%$ of galaxy have a preference for NFW and only $8\%$ of galaxies have strong evidence for NFW.  Once again, we must keep in mind that in certain regions of the parameter space, we expect DC14 to be similar to NFW and thus we expect there to be many galaxies in our sample that have no preference, which in this case is $26\%$ of the entire sample.  In general, the sample tends to favour DC14 much more than NFW and it is clear from this statistic as well as the others provided that for the runs without $\Lambda$CDM priors, the DC14 model is much preferred over NFW.

For the runs with the $\Lambda$CDM priors, we come to a similar conclusion.  We find that $62\%$ of galaxies have a preference for DC14 over NFW and $46\%$ of galaxies have strong evidence in favour of DC14 compared to NFW (see Figure~\ref{BIC}).  Similarly to the runs without $\Lambda$CDM priors, $15\%$ of galaxies prefer NFW while $9\%$ exhibit strong evidence for NFW.  These results are very similar to what we found with the $\Delta$BIC where across the sample, the DC14 model is in general favoured compared to NFW.  Compiling both the frequentist and Bayesian model selection analysis for the runs both with and without $\Lambda$CDM priors, there is compelling and definitive evidence from our study that the DC14 model is preferred.

\subsection{Alternative choice of $\Lambda$CDM priors}
There are systematic uncertainties in the choice of M$_*$/L, mass-concentration, and abundance matching priors and therefore we have run additional experiments to determine how much these affect the model.  Instead of running the full MCMC chains, we take a computationally cheaper approach by simply attempting to maximize the likelihood of the rotation curve fits combined with the priors.  To check for consistency, we have run our maximum likelihood model using the same priors as in the main text and confirmed that it results in a $\chi^2_{\nu}$ CDF consistent with the full MCMC approach.  We have run each minimization from 10 different starting locations in an attempt to avoid local minima.  Because the results here agree with the results from the full MCMC, we are confident that both methods are reaching global minima and this represents another check that the MCMC runs are converged.

\begin{figure}
\centerline{\includegraphics[scale=0.55]{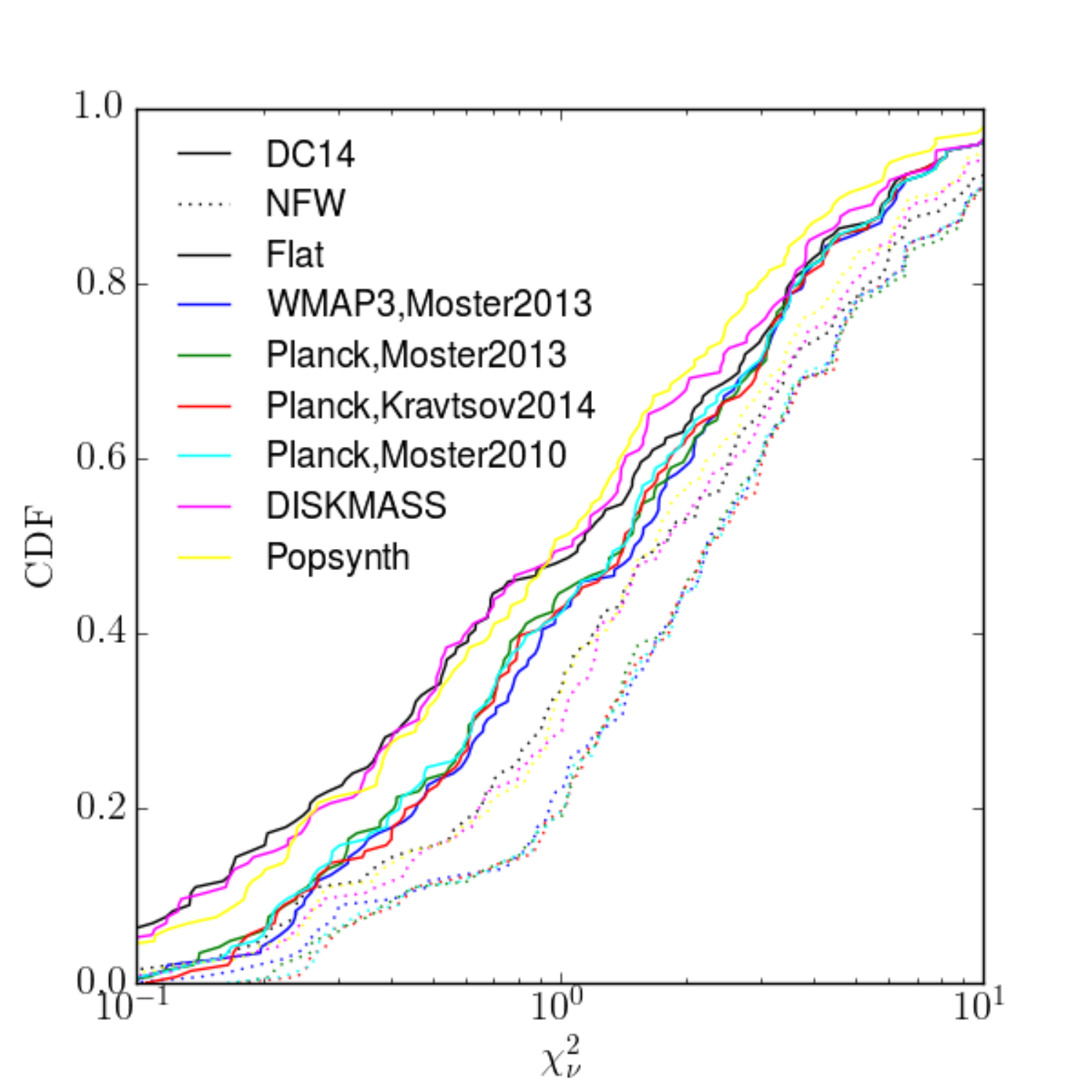}}
\caption{$\chi^2_{\nu}$ CDF comparison where we adopt different combinations of priors as listed in Table~\ref{tab1}.  None of these alternative models disagree with our fiducial set of priors, which indicates that our results are robust to the systematic differences in the derivation of these relations.}
\label{x2multip}
\end{figure}

First, the normalization of the M$_{*}$/L at 3.6$\mu$m is debated \citep{Martinsson2013, Meidt2014, Schombert2014b, McGaugh2014, McGaugh2015}. Different techniques consistently find a small scatter around the mean value of M$_{*}$/L, but the absolute mean value of M$_{*}$/L remains uncertain by a factor of $\sim$2. Our fiducial prior on M$_{*}$/L is flat and adopts a reasonable range of [0.3-0.8], which encompasses different estimates of M$_{*}$/L at 3.6$\mu$ using stellar population synthesis models \citep{Meidt2014,McGaugh2014, Schombert2014b}.  To test whether this has any effect on our conclusions, we refit all galaxies twice using two different Gaussian priors: one with $\rm{M_{*}/L}=0.47\pm0.1\ \text{dex}$ (Popsynth) as given by the self-consistent population synthesis models \citep{McGaugh2014} and one with $\rm{M_{*}/L}=0.24\pm0.05$ as given by the DISKMASS survey \citep{Martinsson2013}, after converting their value from K-band to 3.6 $\mu$m. In these two cases, the M$_{*}$/L may take any value in the wide range [0.1-1.2] but is much more biased towards a specific value. Regardless of the adopted M$_{*}$/L prior, we find no major differences in the quality of the rotation curve fits (See Table~\ref{tab1} and Figure~\ref{x2multip}).  This happens because the difference between DC14 and NFW is observed in low-mass galaxies, where the DM contribution strongly dominates over the baryonic contribution, which is actually often dominated by the gas contribution. 

\begin{table*}
\centering
\begin{tabular}{@{}lccccc@{}}
\hline
M$_*$/L Prior & Mass-concentration & Abundance Matching & colour & $\chi^2_{\nu}$ DC14 & $\chi^2_{\nu}$ NFW \\
\hline
$[0.3-0.8]$ & None & None & black & 1.11 & 1.69 \\
Popsynth & None & None & yellow & 0.97 & 1.69 \\
DISKMASS & None & None & magenta & 1.07 & 1.73 \\
$[0.3-0.8]$ & \cite{Dutton2014} & \cite{Moster2013} & green & 1.41 & 2.23 \\
$[0.3-0.8]$ & \cite{Dutton2014} & \cite{Kravtsov2014} & red & 1.42 & 2.25 \\
$[0.3-0.8]$ & \cite{Dutton2014} & \cite{Moster2010} & cyan & 1.42 & 2.22 \\
$[0.3-0.8]$ & WMAP3$^a$ & \cite{Moster2013} & blue & 1.50 & 2.23 \\
\hline
\end{tabular}
\caption[caption]{List of different runs along with the associated priors and median $\chi^2_{\nu}$.  The colours refer to the lines in Figure~\ref{x2multip}.\\\hspace{\textwidth} $^a$ The WMAP3 mass concentration prior is as given in \protect{\cite{Maccio2008}}.}
\label{tab1}
\end{table*}

\begin{figure}
\centerline{\includegraphics[scale=0.55]{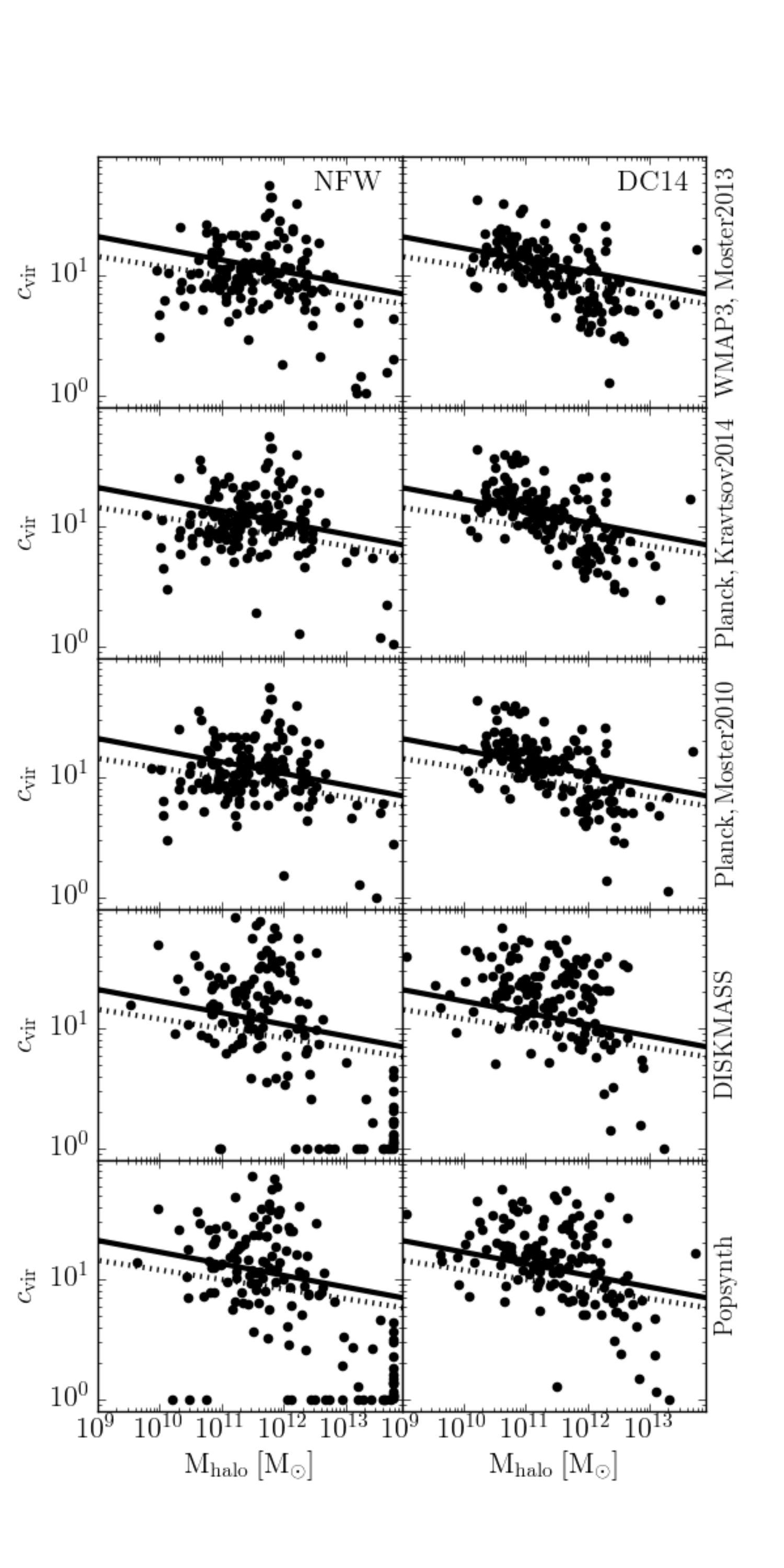}}
\caption{Mass-concentration relations for different choices of priors.  NFW and DC14 halo fits are shown in the left- and right-hand panels, respectively.  The Planck relation from \protect\cite{Dutton2014} is shown as the solid black line, while the WMAP3 relation from \protect\cite{Maccio2008} is shown as the dotted line.  The imposed priors are listed on the right of each row and can be found in Table~\protect\ref{tab1}.}
\label{MCmultip}
\end{figure}

While the fits to the rotation curves are comparable between these two M$_{*}$/L priors, we do see a difference in their ability to adhere to the mass-concentration relations.  In Figure~\ref{MCmultip}, we plot the mass-concentration relations for our maximum likelihood halo parameters using these new sets of priors.  We can see that the DISKMASS prior tends to cause the parameters to scatter higher than the relations for the DC14 model.  The normalization of the DISKMASS M$_{*}$/L is in disagreement with that of other studies \citep{Meidt2014,McGaugh2014,Schombert2014b} and this lower normalization increases the halo concentration as more DM mass will need to be packed into the centres of the galaxies in order to make up for the decrease in stellar mass due to a lower M$_{*}$/L.  The normalization we find for the Popsynth halo fits are fairly identical to what we have found in our full MCMC.  Nevertheless, we do note that there is additional scatter when applying the M$_{*}$/L priors in this fashion compared to a log flat prior on M$_{*}$/L in the range [0.3-0.8], simply because of the additional freedom.  We must be cautious in allowing too much freedom in M$_{*}$/L, as this will become inconsistent with the scatter expected from population synthesis models.  More importantly, the stellar masses may scatter so much that a tight baryonic Tully-Fisher relation may no longer be obtained.  The fits to the NFW haloes for these models also show an increased scatter compared to our original fits.  Furthermore, many galaxies still continue to show ``core-cusp" issues as many of the maximum likelihood halo parameters remain pinned to our lower limit on concentration and upper limit on viral velocity.

\begin{figure}
\centerline{\includegraphics[scale=0.55]{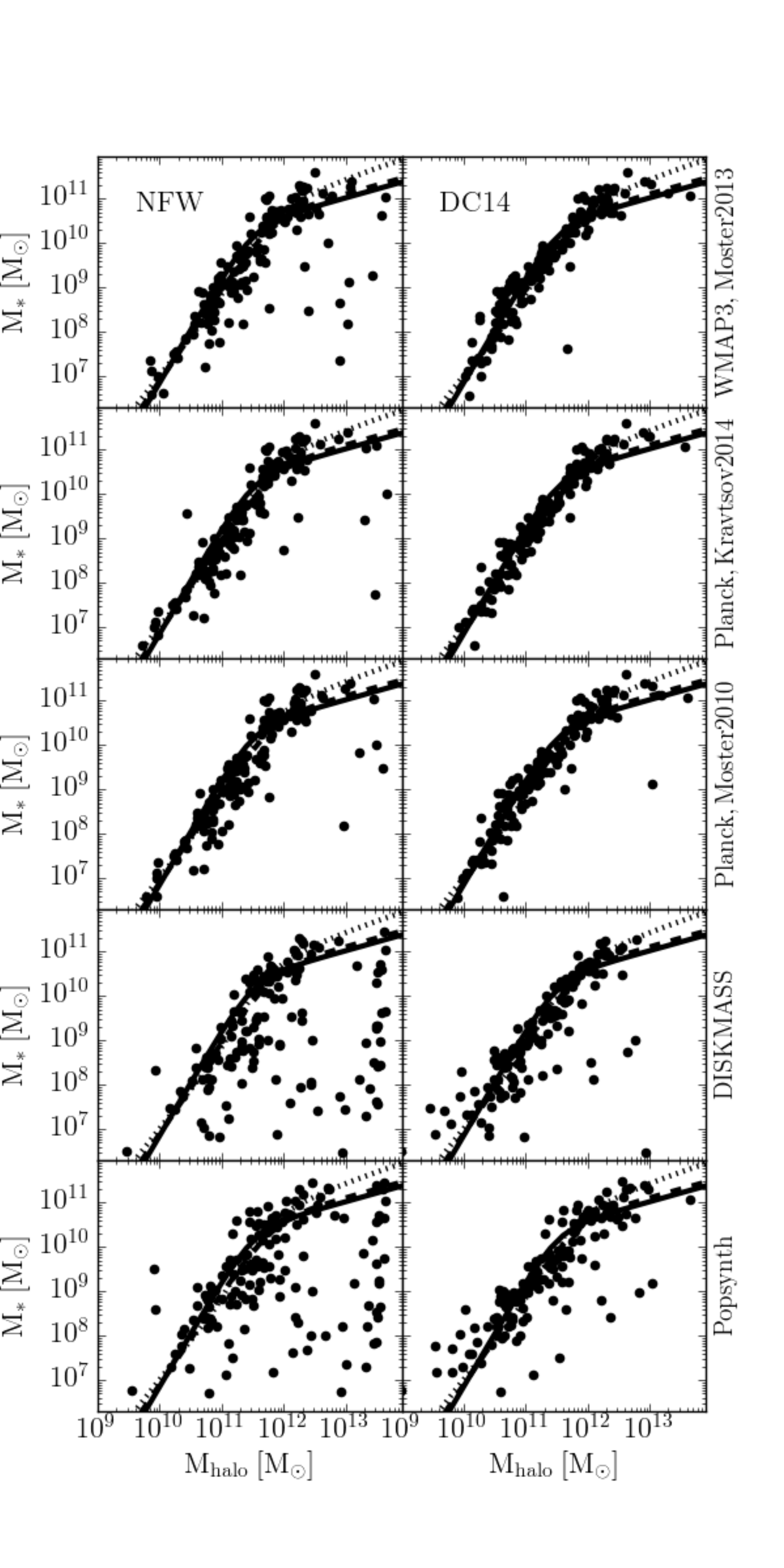}}
\caption{M$_*$-M$_{\text{halo}}$ relations for different choices of priors.  NFW and DC14 halo fits are shown in the left- and right-hand panels, respectively.  The \protect\cite{Moster2013}, \protect\cite{Moster2010}, and \protect\cite{Kravtsov2014} relations are shown as the solid, dashed, and dotted lines, respectively.  The imposed priors are listed on the right of each row and can be found in Table~\protect\ref{tab1}.}
\label{AMmultip}
\end{figure}

Contrary to the mass-concentration relation, we find no discernible difference between the Popsynth and DISKMASS M$_{*}$/L priors in the M$_{*}$-M$_{\text{halo}}$ relation (see Figure~\ref{AMmultip}).  The DC14 model halo parameters continue to cluster around the expected relations, although we do see a few instances of points falling low.  Part of this certainly has to do with the fact that we expect many of these posteriors to be multimodal.  In some cases, the non-cosmologically motivated mode may have a lower $\chi^2_{\nu}$ and our simple maximum likelihood estimate cannot deal with this in the same way as we have done for the full MCMC fits.  As mentioned before, we do expect the scatter to increase to some degree simply because there is more freedom in M$_*$/L.  Similarly, there is no clear indication that the NFW halo model fits cluster around the expected relations with these alternative M$_{*}$/L priors and the scatter is significantly larger than what we see for the DC14 model.

There is some debate in the literature about the functional form of the ${\rm M_{*}-M_{\text{halo}}}$ relation. Thus, we have refitted all galaxies using a different abundance matching relation \citep{Kravtsov2014}, which deviates strongly at the high-mass end from our fiducial ${\rm M_{*}-M_{\text{halo}}}$ relation \citep{Moster2013}. We find that this has no discernible effect on our conclusions (see Table~\ref{tab1} and Figure~\ref{x2multip}).  Furthermore we also attempted the same exercise adopting the ${\rm M_{*}-M_{\text{halo}}}$ from \cite{Moster2010} that uses a slightly different cosmology and also find no effect on the results (See Table~\ref{tab1} and Figure~\ref{x2multip}).  It can be seen in Figure~\ref{MCmultip} that neither of these alternative ${\rm M_{*}-M_{\text{halo}}}$ relations significantly affect our estimate of the mass-concentration relation.  These three ${\rm M_{*}-M_{\text{halo}}}$ relations agree well at lower masses and the only major difference occurs at ${\rm M\gtrsim 4\times10^{11}\ M_{\odot}}$.  In this regime the DC14 halo model fits seem to cluster slightly closer to the relation from \cite{Kravtsov2014}.  This is true also for our NFW model fits but unsurprising since we can see from Figure~\ref{quadlcdm} that for our full MCMC fits, haloes of this mass tend to fall on the higher side of the \cite{Moster2013} relation, but still remain consistent within two standard deviations.

Finally, we can question which is the best mass-concentration relation to use.  We have performed the same exercise, adopting the mass-concentration relation from \cite{Maccio2008} which is for WMAP3 cosmology and find very similar results as before.  There is an extremely marginal increase in $\chi^2_{\nu}$ from 1.41 to 1.50 when switching to the WMAP3 cosmology, and it would be difficult to differentiate between the two models based on this statistic alone.  When applying the WMAP3 prior, the concentrations of the halo parameters decrease to become more consistent with the lower normalization in the mass-concentration relation of WMAP3 compared to Planck.  This, however, does not significantly affect the quality of the rotation curve fits for the DC14 or the NFW halo models.  Likewise, there is no tension between the  ${\rm M_{*}-M_{\text{halo}}}$ relations and the DC14 model halo parameters when the WMAP3 prior is imposed.  The bottom line is that the median $\chi^2_{\nu}$ for the NFW fits without imposing any of the different $\Lambda$CDM priors are always worse than any of the fits with the DC14 model, even with the $\Lambda$CDM priors imposed.

\subsection{Caveats}
As we have mentioned, rotation curve data and galaxy mass models are far from perfect as there are uncertainties in distance and non-circular motions and here we only consider smooth spherical halo profiles.  Observationally the SPARC data set is a major step forward as it was analysed in a homogenous fashion.  Theoretically, the DC14 halo profile considers processes in galaxy formation beyond gravitational collapse \citep{DC2014}.  Comparing our maximum posterior halo parameters to the ${\rm M_{*}-M_{\text{halo}}}$ and mass-concentration relations demonstrates the consistency between the DC14 model, the rotation curves, and expectations from cosmology.  However, certain systematics persist that may inherently prevent a perfect agreement.  

The stellar masses were computed in a very different fashion in our modelling compared to what was done for the ${\rm M_{*}-M_{\text{halo}}}$ relations \citep{Moster2013}.  This can, in principle, systematically bias both relations.  Furthermore, the ${\rm M_{*}-M_{\text{halo}}}$ relation considers all galaxies and not only late-type galaxies that exhibit a high degree of rotational symmetry.  The slope of the ${\rm M_{*}-M_{\text{halo}}}$ relation may be different depending on the type of galaxy.  This systematic appears in the mass-concentration relation as well, which was derived from a DM-only simulation and thus cannot isolate the relation for only symmetric late-type galaxies \citep{Dutton2014}.  Additionally, we have assumed a constant M$_{*}$/L across the entire galaxy while the bulge region may have a higher M$_{*}$/L compared to the disk.  This can change how the inner regions of the most massive galaxies in our sample are fit, affecting halo concentrations.  Finally, the DC14 model does well in predicting the general trends in the expected $\alpha,\ \beta,\ \&\ \gamma$, however this is clearly a mean relation that will not fit every galaxy perfectly \citep{DC2014}.  The history of an individual galaxy is likely to matter and one can expect that a galaxy that undergoes a single starburst at high redshift will have a different halo profile to a galaxy of a similar mass that has had a much more tempered star formation history.  The impulsiveness of the starburst can change how the halo responds \citep{Pontzen2012}.  For these reasons, we expect a certain amount of disagreement in the halo fitting and the relations we compare to.  However, over the whole sample, the DC14 model does remarkably well suggesting that some of these other effects in galaxy formation may be of second order in determining the halo profile parameters.

\section{Conclusions}
\label{DandC}
In this paper, we have tested two different DM halo profiles: the universal NFW profile obtained from DM-only simulations, and the M$_*$/M$_{\text{halo}}$-dependent DC14 profile obtained from hydrodynamic simulations. We fit 147 rotation curves from the SPARC data set using an MCMC technique. We find that the DC14 profile provides better fits to the rotation curves than NFW. This result holds for several different choices of priors on M$_*$/L as well as $\Lambda$CDM scaling relations.

The DC14 halo model can provide good fits to the observed rotation curves of galaxies over a large range in luminosity and surface-brightness. This modification of the DM-only NFW form is restricted to the inner regions of galaxies, yet reproduces more global relations like M$_*$-M$_{\text{halo}}$.  Since the observed rotation curves do not probe out to the virial radius of the halo, they can only constrain the behaviour of the density profile in the inner portions of galaxies.  This can lead to large uncertainties on M$_{\text{halo}}$ (see Figure~\ref{quadflat}).  However, the maximum posterior halo parameters for the DC14 rotation curve adhere well to the M$_*$-M$_{\text{halo}}$ relation which suggests that the DC14 model has the correct density profile at large as well as small radii.

The cores observed in dwarf and LSB galaxies supports the notion that baryonic processes result in the net expansion of primordial DM density profiles.  It appears that the feedback prescription employed to produce the DC14 halo model is on the right track to modify primordial DM haloes to be consistent with observations.  In particular, the magnitude, timing, and spatial distribution of outflows lead to the appropriate halo transformation required to explain observed rotation curves.  This is an important step forward in galaxy formation modelling. Optimistically, it resolves a hotly debated discrepancy over two decades old. 

It remains to be seen whether real galaxies actually experience the rapid, repeated outflows of large quantities of gas that are required to transform cusps into cores.  In the Local Universe, starburst dwarf galaxies often show outflows of ionized gas \citep{Martin1998,Schwartz2004}, but the mass involved in these outflows seems to be a very small fraction of total gas mass \citep{Lelli2014}.  The situation may be different in the early Universe \citep{Erb2015}, but unfortunately the observations of high-redshift galaxies are more difficult to interpret. Further observations of both local and high redshift outflows will be critical in determining whether such processes can indeed reconcile the discrepancy between cuspy DM haloes and the slowly rising rotation curves observed in real galaxies.

\section*{Acknowledgements}
H.K thanks Simon Gibbons, Angus Williams, and Sergey Koposov for useful discussions as well as all others who have commented on the manuscript.  We thank the anonymous referee for reviewing and improving the manuscript.  H.K. is supported by and thankful to Foundation Boustany, the Isaac Newton Studentship, and the Cambridge Overseas Trust.  ADC is supported by the DARK independent fellowship program and the Carlsberg foundation. The Dark Cosmology Centre is funded by the Danish National Research Foundation.  CBB thanks the MICINN (Spain) for the financial support through the MINECO grant AYA2012-31101 and  the Ramon y Cajal program.  This publication was made possible through the support of the John Templeton Foundation.  The opinions expressed in this paper are those of the author and do not necessary reflect the views of the John Templeton Foundation.

\bibliographystyle{./mn2e}
\bibliography{./SPARC_DC14_clean.bib}

\appendix
\section{Selected Rotation Curves and Posteriors}
\label{rcplots}
Here we discuss a few select cases from the SPARC sample to show qualitatively how the posteriors and rotation curve fits look for different masses, priors, confidence interval size, and multimodal posteriors.  We have attempted to isolate certain peculiar cases to address many of the claims made in the main text.  We show examples with both good and bad fits to the rotation curves in terms of $\chi^2_{\nu}$.

\begin{figure*}
\centerline{\includegraphics[scale=0.5]{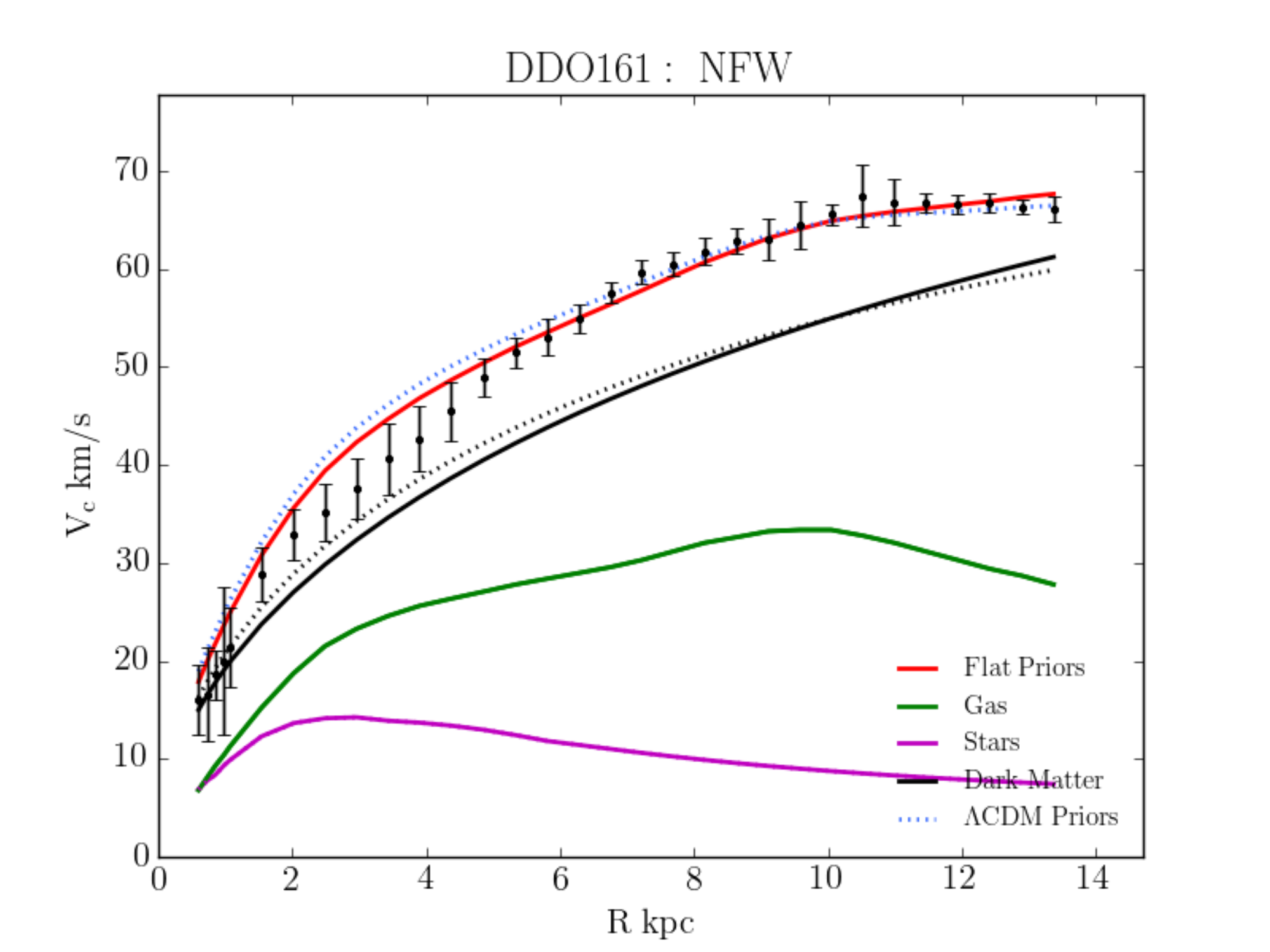}\includegraphics[scale=0.5]{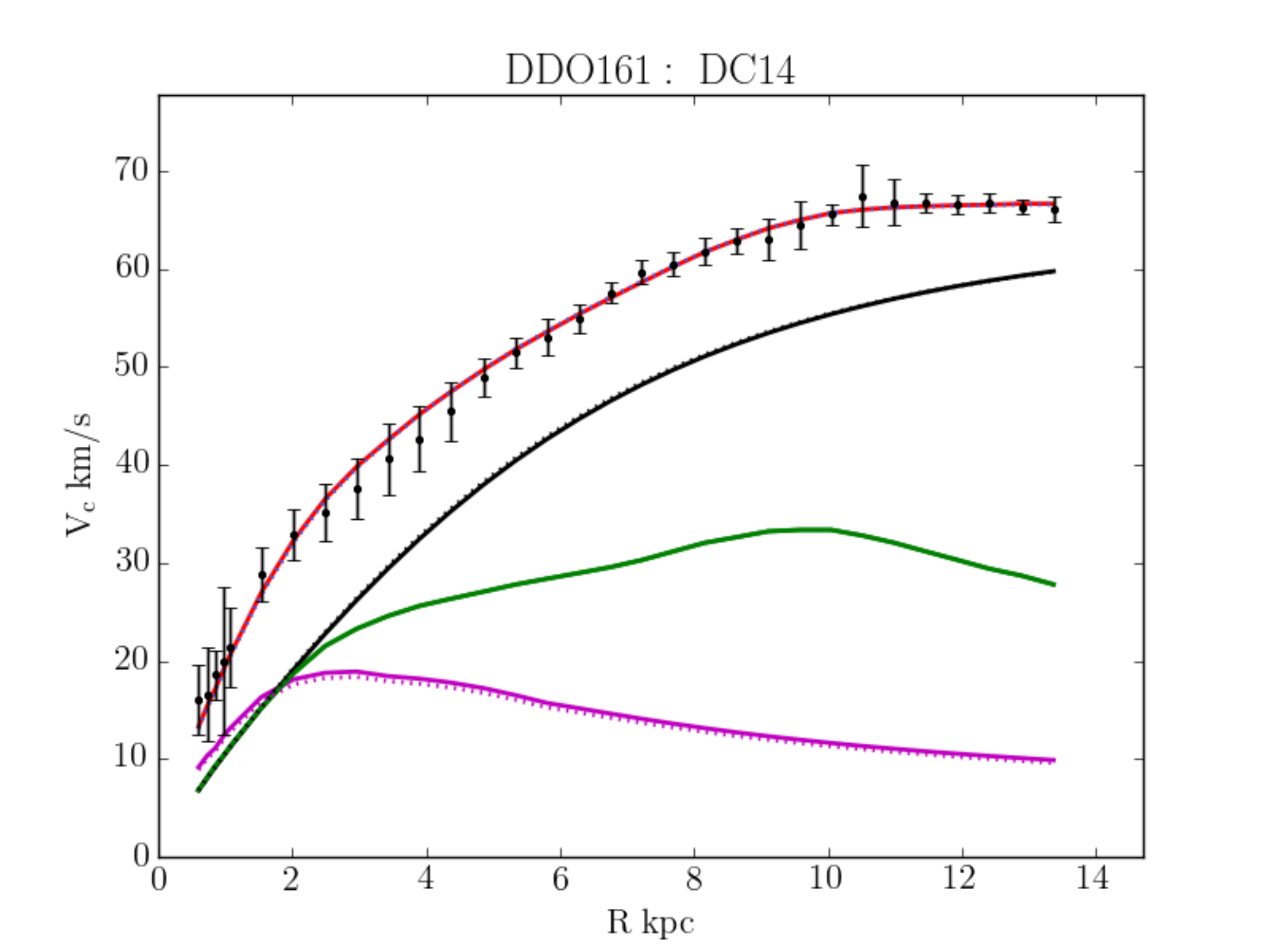}}
\centerline{\includegraphics[scale=0.5]{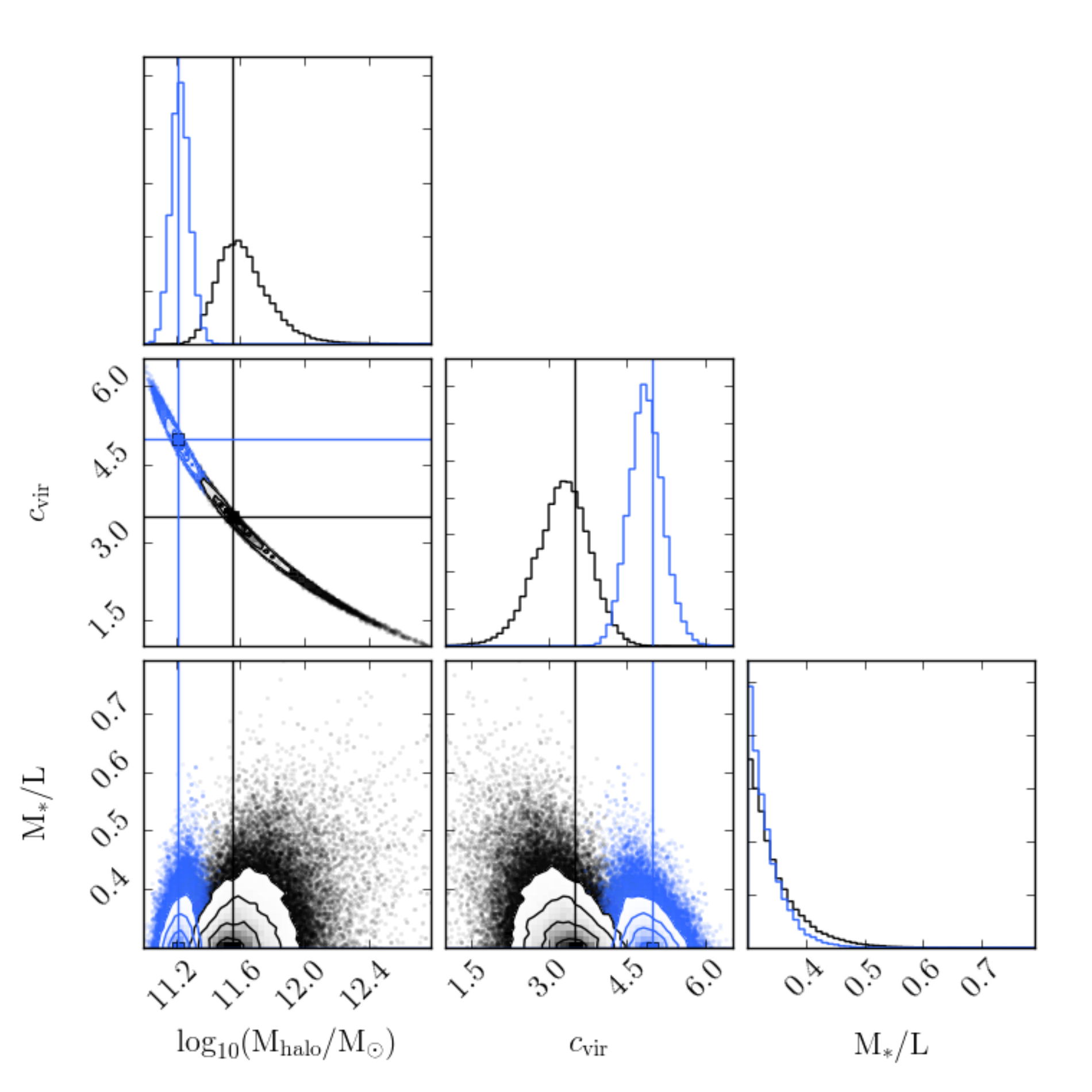}\includegraphics[scale=0.5]{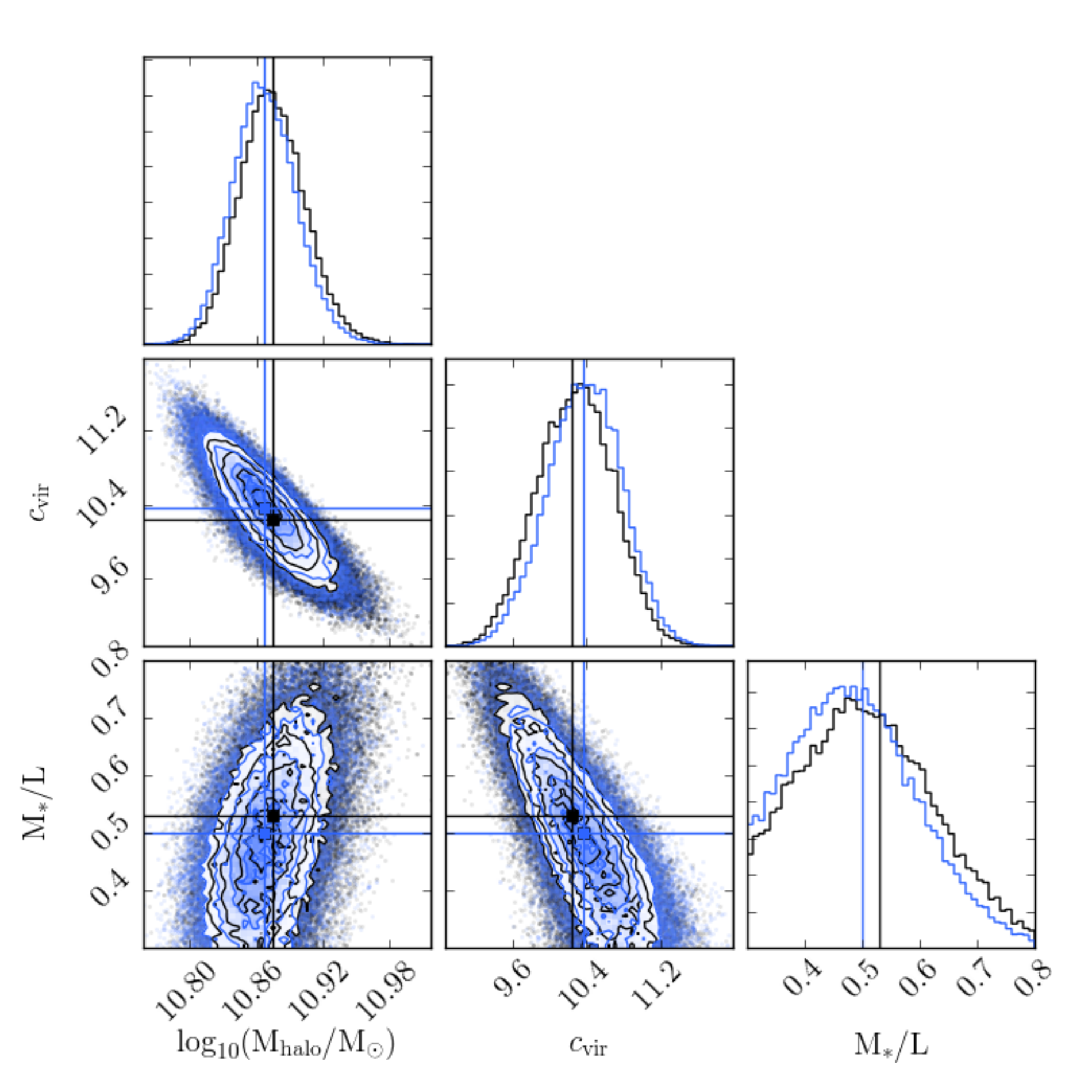}}
\caption{{\bf DDO161:} (Top) Rotation curve fits for the maximum posterior halo models for NFW (Left) and DC14 (Right).  The solid red lines show the fits with the fiducial flat priors while the dotted blue lines show the fits with the $\Lambda$CDM priors imposed.  The gas, stellar, and DM contributions are shown in green, magenta, and black respectively.  For these quantities, the dotted lines show the values for the $\Lambda$CDM priors while solid lines show the values for the fiducial flat priors.  The gas contribution is always the same regardless of the prior.  Secondary modes, when present, are shown as the more translucent lines.  (Bottom)  Posterior distributions for the fitting parameters for the NFW model (left), and the DC14 model (Right).  The black and blue lines and contours show the parameter space and 1D marginalized posterior for the fits with the fiducial flat priors and $\Lambda$CDM priors, respetively, consistent with the top panel.  The blue point shows the maximum posterior parameters for the fits with the $\Lambda$CDM priors while the black point shows the maximum posterior point for the fits with the fiducial flat priors.  If secondary modes are present, the maximum posterior points are shown in yellow and green for the fits without and with $\Lambda$CDM priors, respetively.  The corner plots were made with the open-source software {\small corner.py} \protect{\citep{corner}}.}
\label{DDO161}
\end{figure*}

{\bf DDO161:} DDO161 is a low-mass galaxy with ${\rm V_{flat}}\sim65$ and in Figure~\ref{DDO161}, we show the posteriors and rotation curve fits.  For the NFW model, the shape of the posterior in the mass-concentration plane is a characteristic crescent.  The $\Lambda$CDM priors have a strong effect on the halo parameters for the NFW model as the concentration shifts towards higher values and the mass decrease.  On the contrary, the DC14 model shows very good agreement between the runs with and without $\Lambda$CDM priors, the posteriors completely overlap.  For this galaxy, we have a very tight constraint on mass and concentration while the fit is insensitive to ${\rm M_*/L}$.  This is because DDO161 is a gas dominated galaxy.  Nevertheless, the NFW is fitting a cusp to a galaxy that prefers a core, ${\rm M_*/L}$ is pinned against the lower limit of the prior for the NFW fit; hence, these trends are common among galaxies of this mass.

Although the NFW model over-predicts the rotation velocity for this galaxy towards the centre, the $\chi^2_{\nu}$ of all of the fits are reasonably good.  The DC14 model has  $\chi^2_{\nu}=0.24$ regardless of the priors while  $\chi^2_{\nu}$ increases from 0.93 to 1.33 when the $\Lambda$CDM priors are imposed.

\begin{figure*}
\centerline{\includegraphics[scale=0.5]{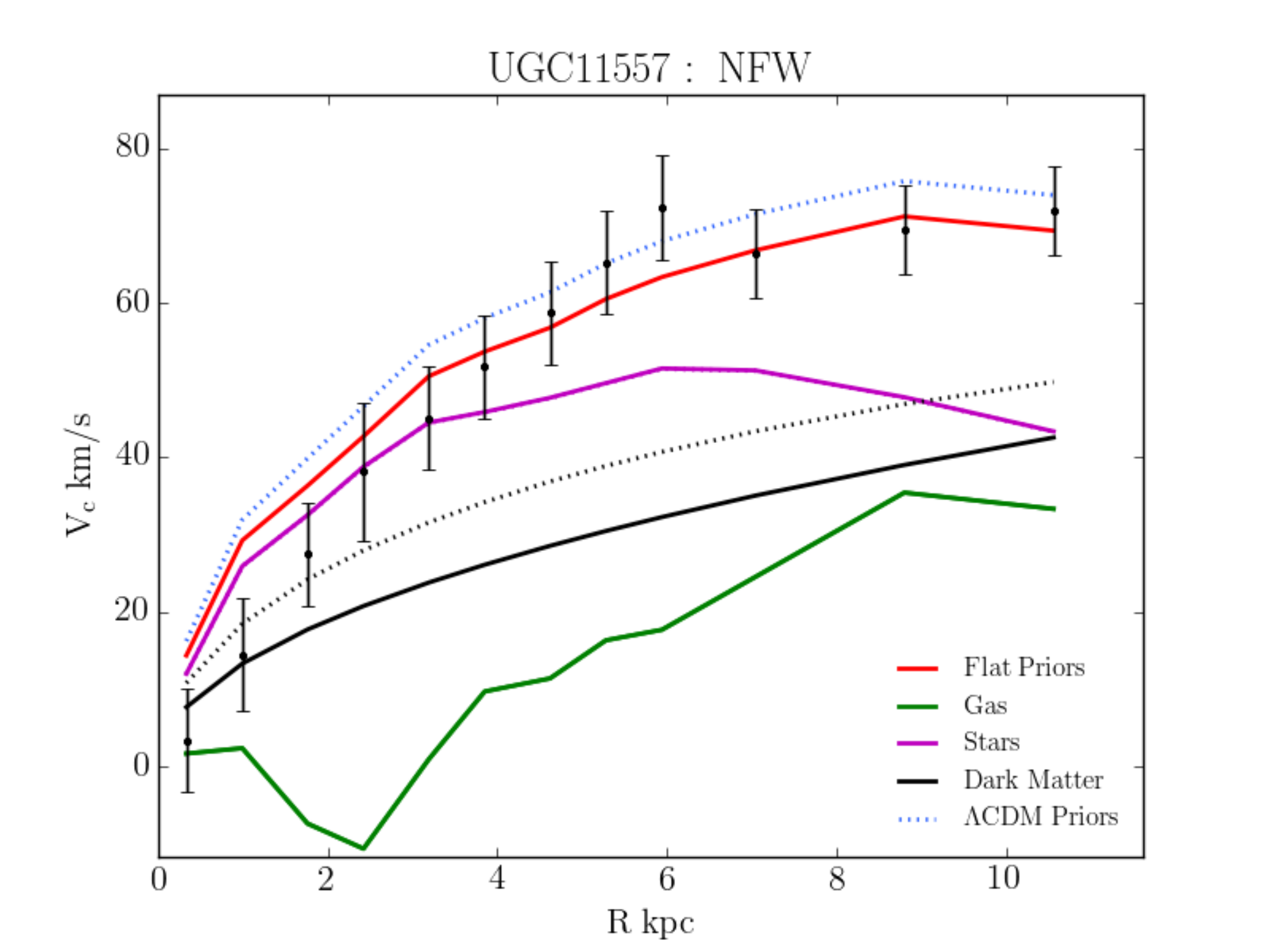}\includegraphics[scale=0.5]{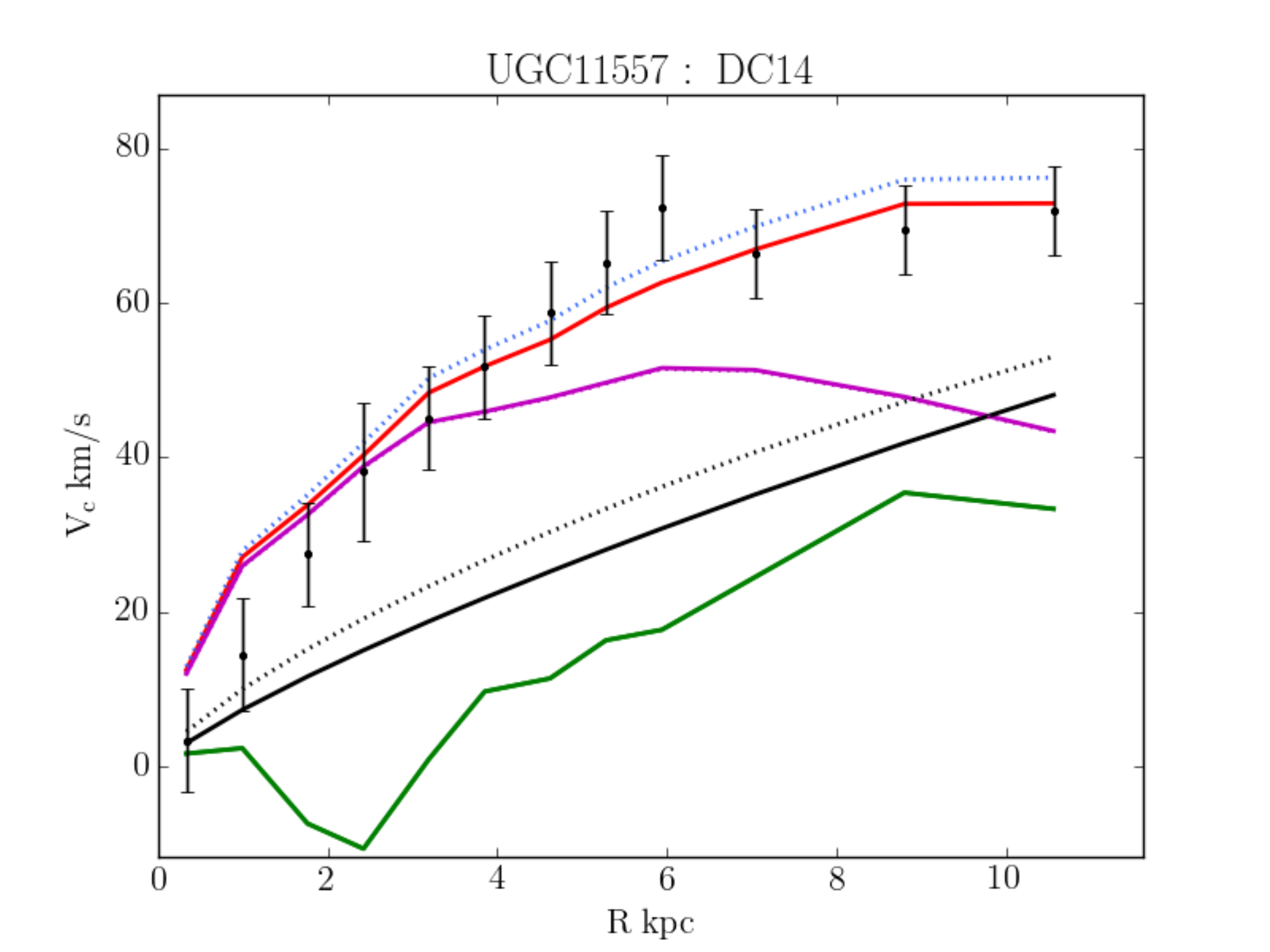}}
\centerline{\includegraphics[scale=0.5]{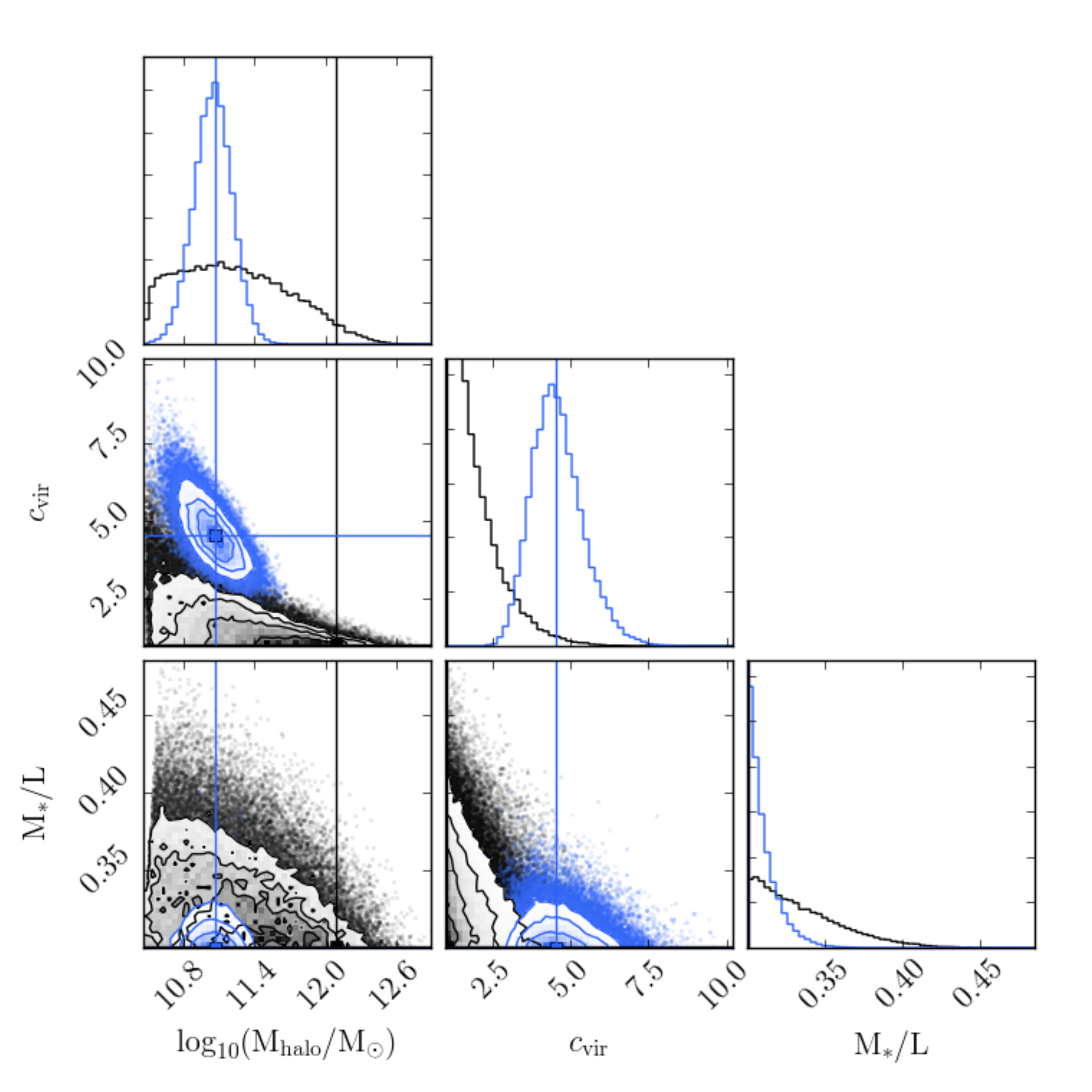}\includegraphics[scale=0.5]{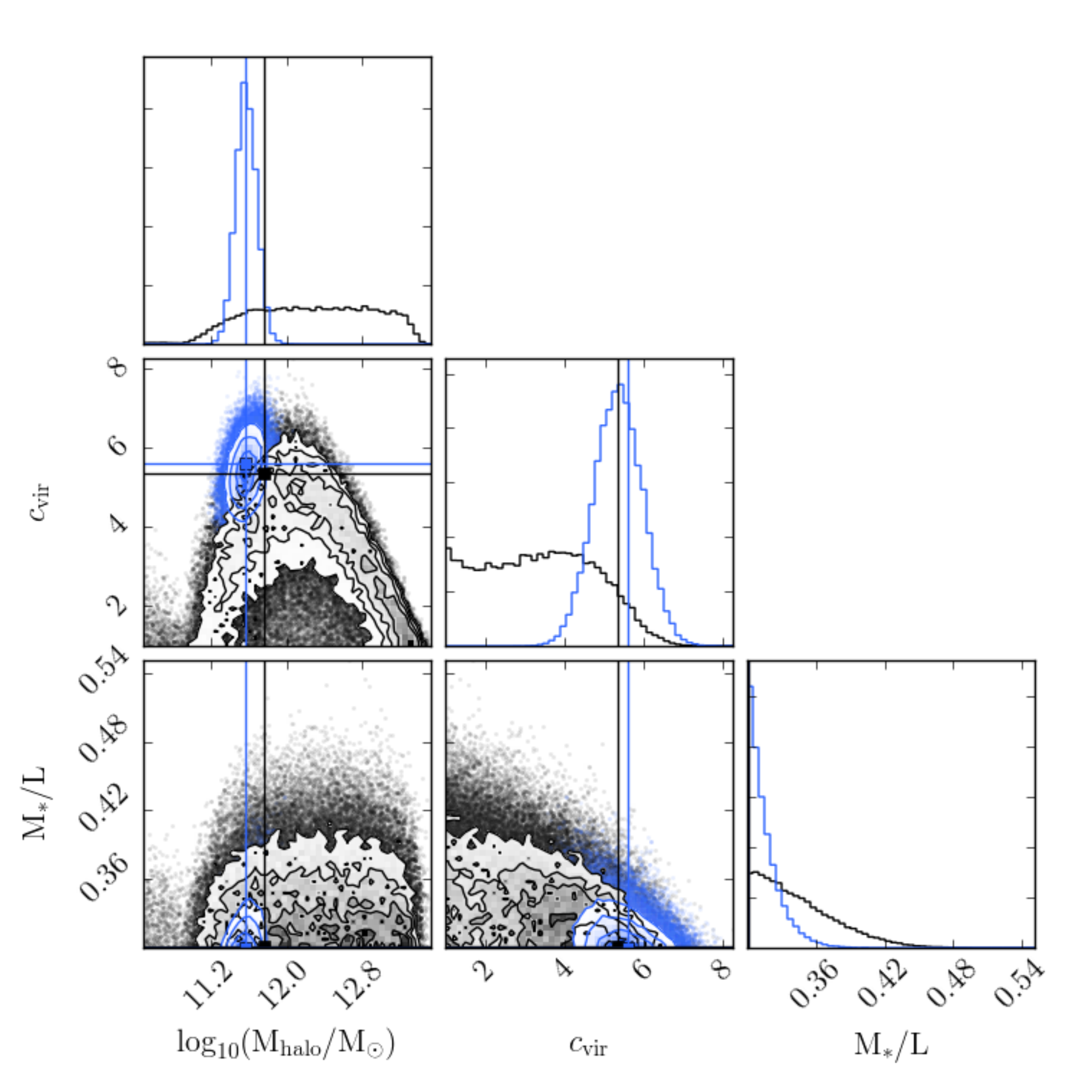}}
\caption{{\bf UGC11557:} See Figure~\ref{DDO161} for caption.}
\label{UGC11557}
\end{figure*}

{\bf UGC11557:}  UGC11557 is another example of a lower mass galaxy but the rotation curve is much more uncertain.  Unlike DDO161, the baryons are dominated by the stars (see Figure~\ref{UGC11557}).  Even at low ${\rm M_*/L}$, the disk is maximal in the inner regions which biases ${\rm M_*/L}$ to lower values for both halo models.  For both the DC14 and NFW models, the 95\% confidence interval on mass is large as the posterior is very wide and it spans almost two orders of magnitude.  For the NFW model, the concentration is pinned to the lower edge of the prior because the rotation curve is maximal in the inner regions.  This does not necessarily have to be the case for the DC14 model as the concentration spans out to reasonable values.  The posterior of the DC14 model in the mass-concentration plane is extremely peculiar, creating an arch shape.  For this reason, the 1D error bars on mass are extremely misleading as at a specific mass there is still only a small range in concentration where the halo fit is valid.

When the $\Lambda$CDM priors are imposed, the posterior distributions look very different for both galaxies.  The posterior distribution for the NFW halo, with the $\Lambda$CDM priors is completely inconsistent with the posterior distribution without the priors.  The maximum posterior halo parameters wildly vary and this causes the $\chi^2_{\nu}$ to increase from 1.53 to 2.42.  The posterior distributions for the DC14 model are also very different with and without $\Lambda$CDM priors.  There is a much tighter range in parameter space allowed when the $\Lambda$CDM priors are imposed, but conveniently, the maximum posterior halo parameters do not change much for this model.  $\chi^2_{\nu}$ increases from 1.19 to 1.39 which would still be considered a reasonable fit.  In this case, the NFW model cannot provide a good fit to the rotation curve while the DC14 model can.

\begin{figure*}
\centerline{\includegraphics[scale=0.5]{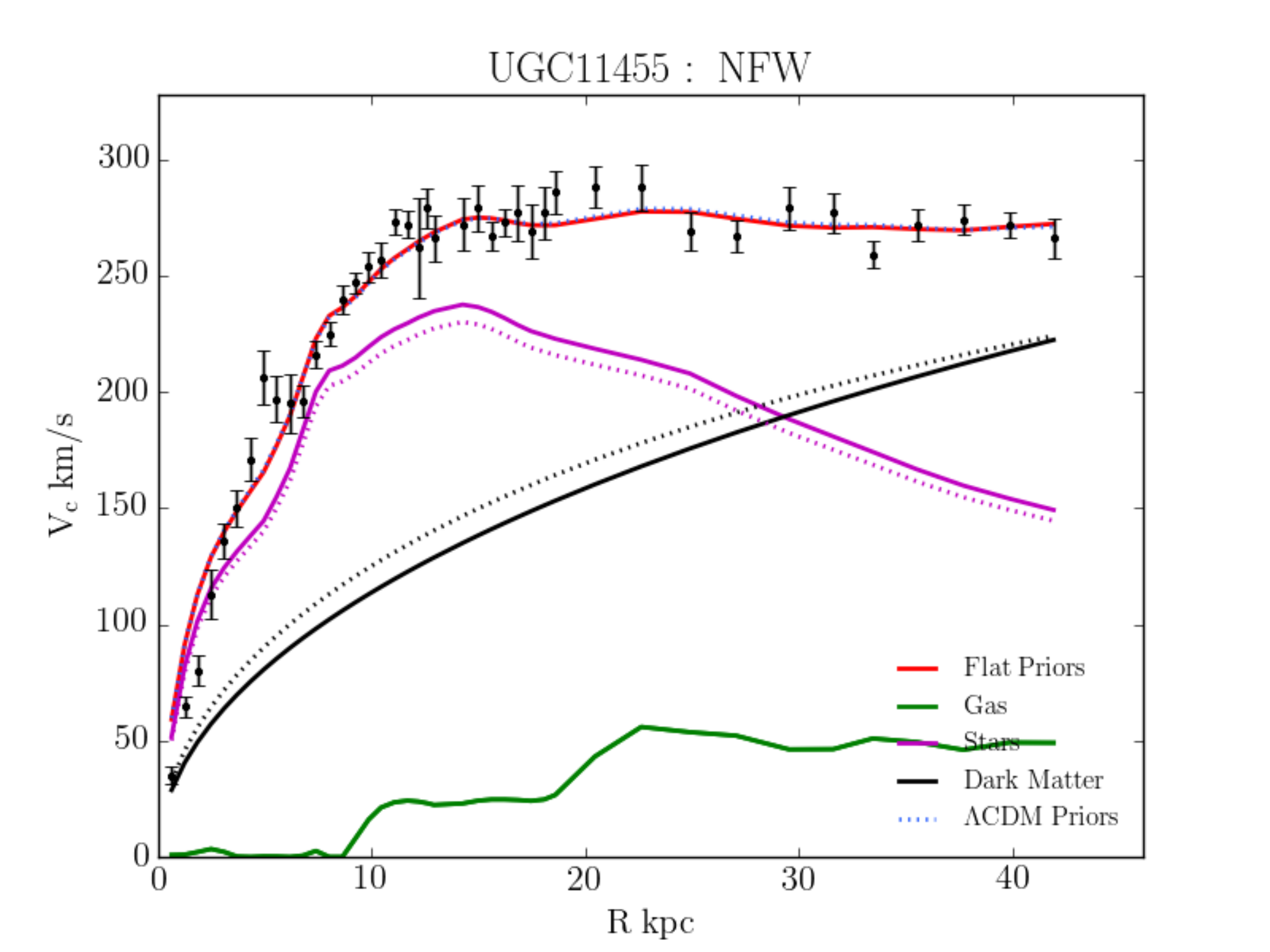}\includegraphics[scale=0.5]{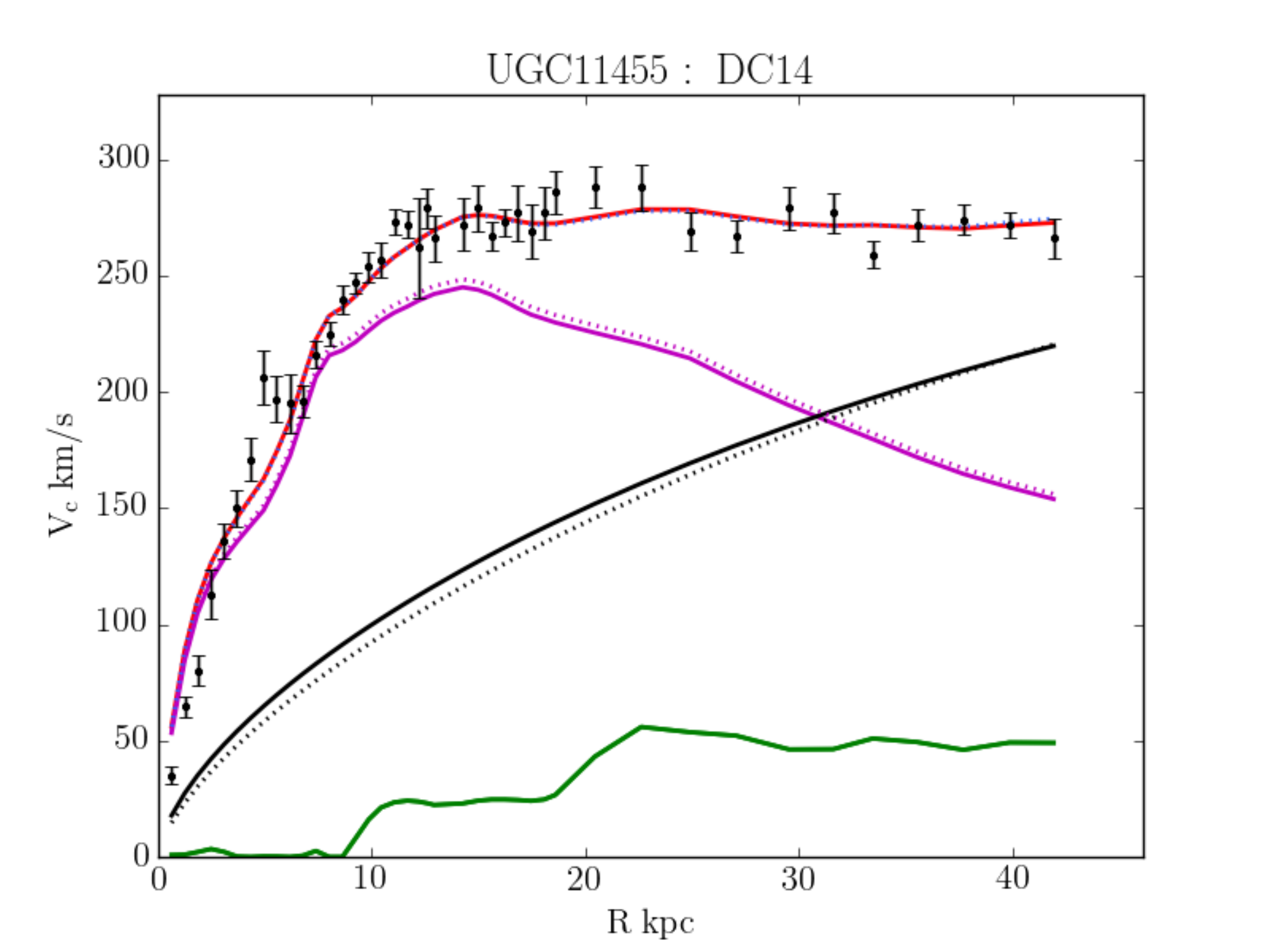}}
\centerline{\includegraphics[scale=0.5]{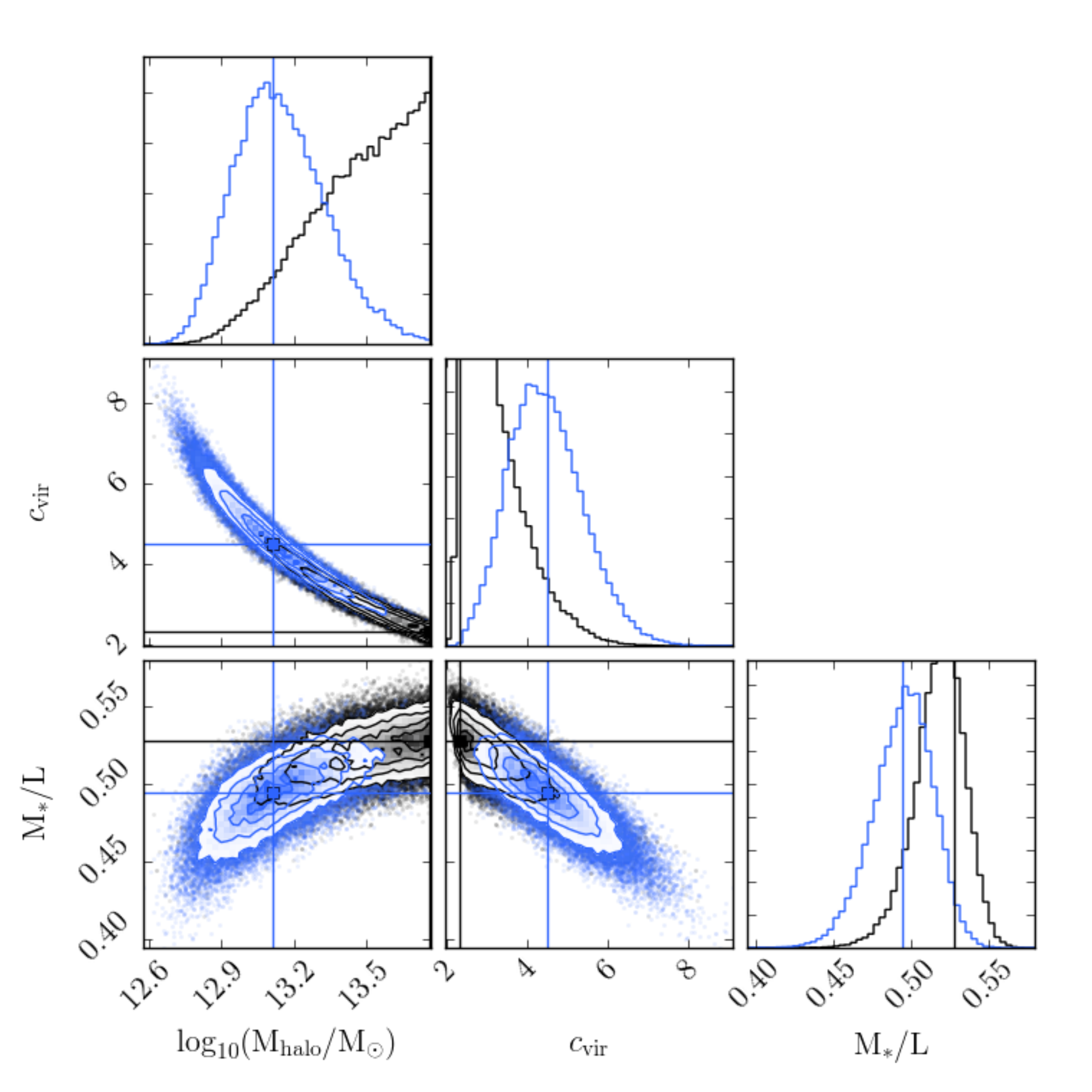}\includegraphics[scale=0.5]{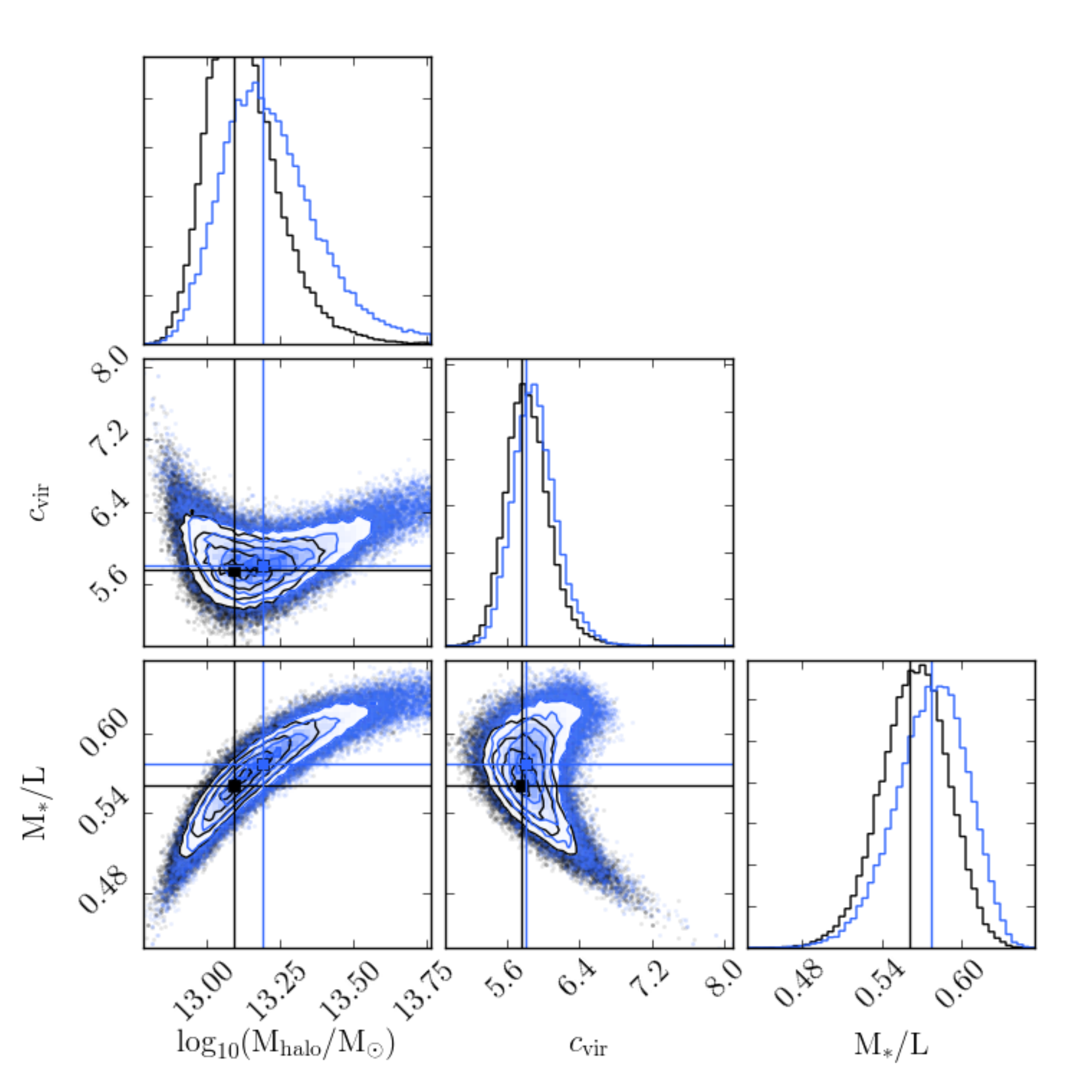}}
\caption{{\bf UGC11455:} See Figure~\ref{DDO161} for caption.}
\label{UGC11455}
\end{figure*}

{\bf UGC11455:} UGC11455 is a high-mass high-surface-brightness galaxy (See Figure~\ref{UGC11455}).  It has a slightly super maximal disk at the smallest values radii but requires a DM component at $r\gtrsim7$~kpc.  For all intents and purposes, one would say that both halo models do a very good job at matching the general shape of this rotation curve.  ${\rm V_{flat}}$ is well predicted and the steep rise at low $r$ is captured by the models (mainly due to the baryonic component).  However, this is an example of a galaxy where the $\chi^2_{\nu}$ value is bad, being $\sim4$ for the DC14 fits and $\sim4.5$ for the NFW fits, while qualitatively one might say this is a good fit.  There are bumps and wiggles in the rotation curve that cannot be fitted with a smooth halo model.  We have shown this galaxy to demonstrate the reasons why higher luminosity galaxies are biased towards higher $\chi^2_{\nu}$ values.  However, we do believe that the halo parameters derived for this galaxy are reasonably representative given the fact that the important qualitative features are well matched by the fits.  There are many galaxies of this ilk in our data set.

Looking at the posterior distributions, we can see in Figure~\ref{UGC11455} that without $\Lambda$CDM priors, the mass of the NFW halo stretches to the high end of the prior while the concentration is pinned to low values.  This is not the case for the DC14 halo, which nicely picks out reasonable parameters when no priors are imposed.  With the $\Lambda$CDM priors imposed, the maximum posterior halo parameters do not change significantly for the DC14 model while once again we see a large difference for NFW.

\begin{figure*}
\centerline{\includegraphics[scale=0.5]{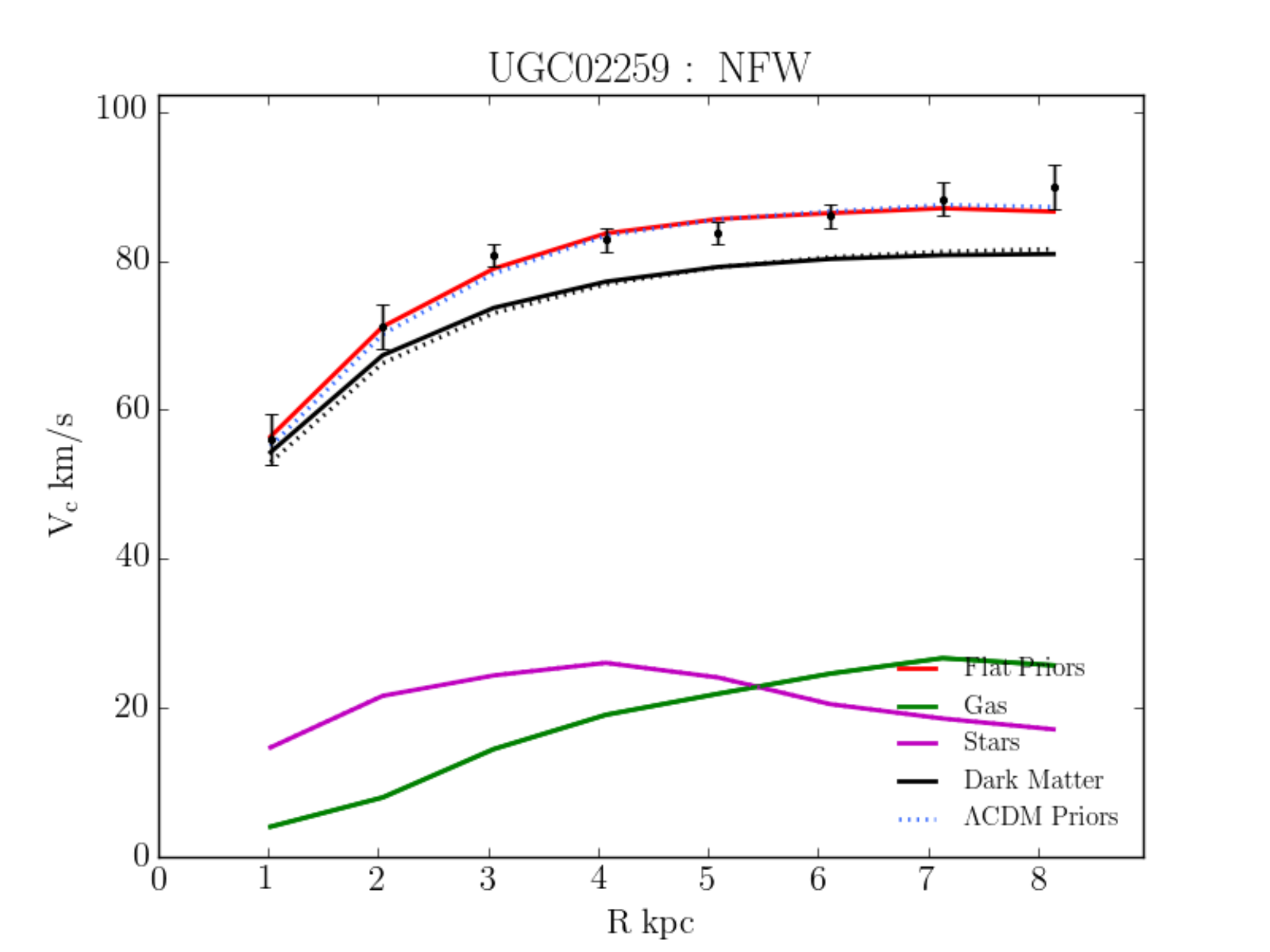}\includegraphics[scale=0.5]{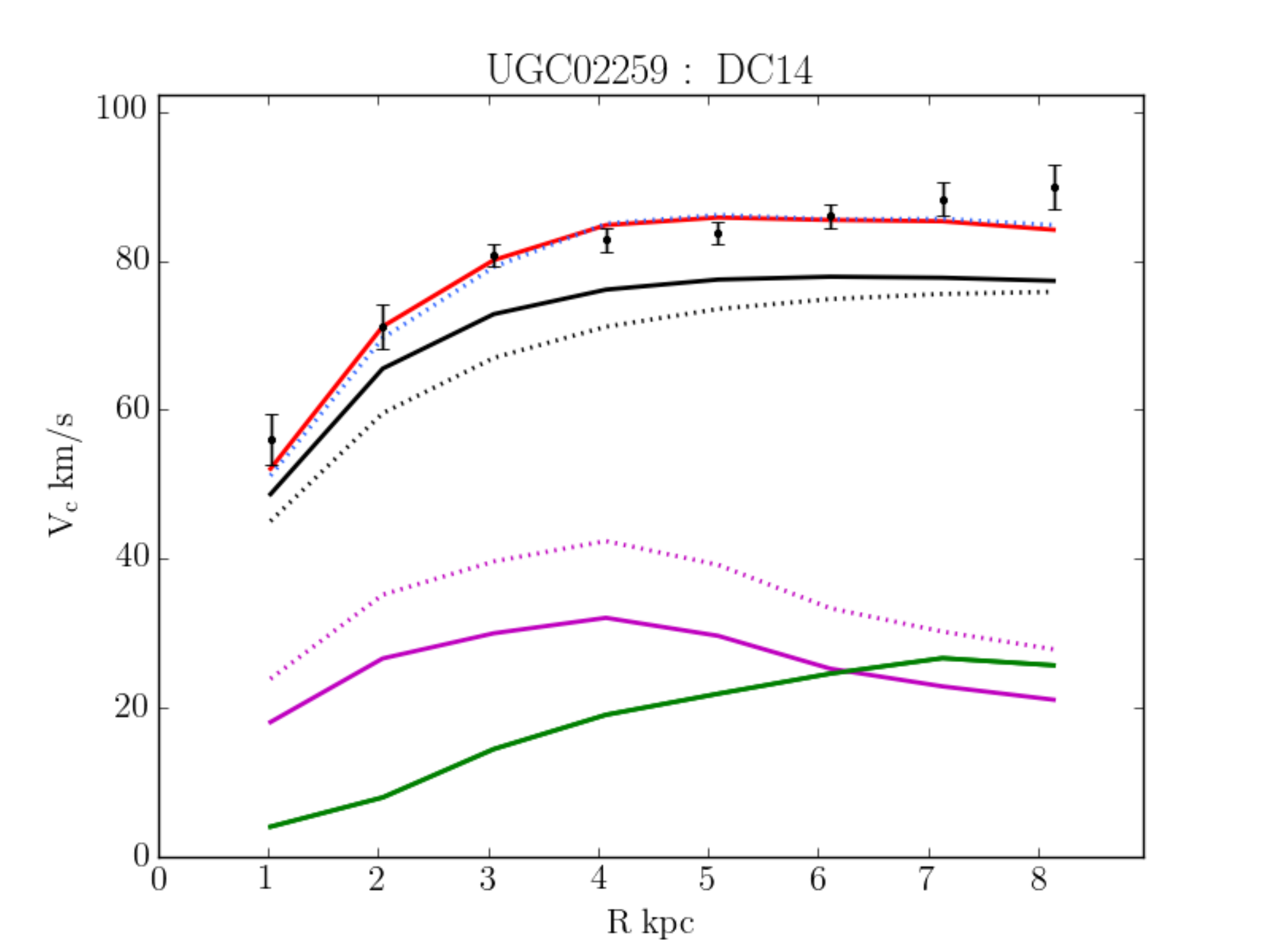}}
\centerline{\includegraphics[scale=0.5]{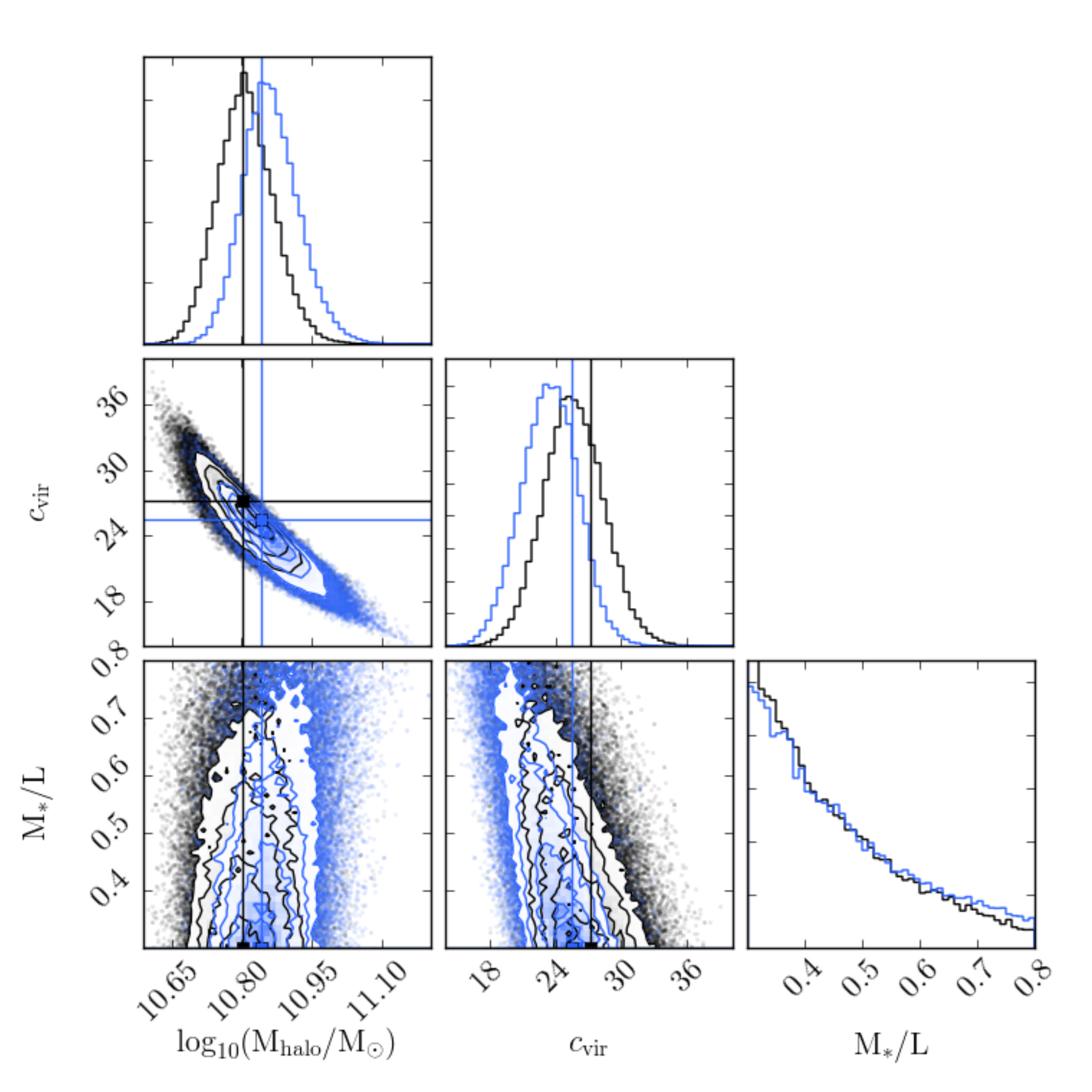}\includegraphics[scale=0.5]{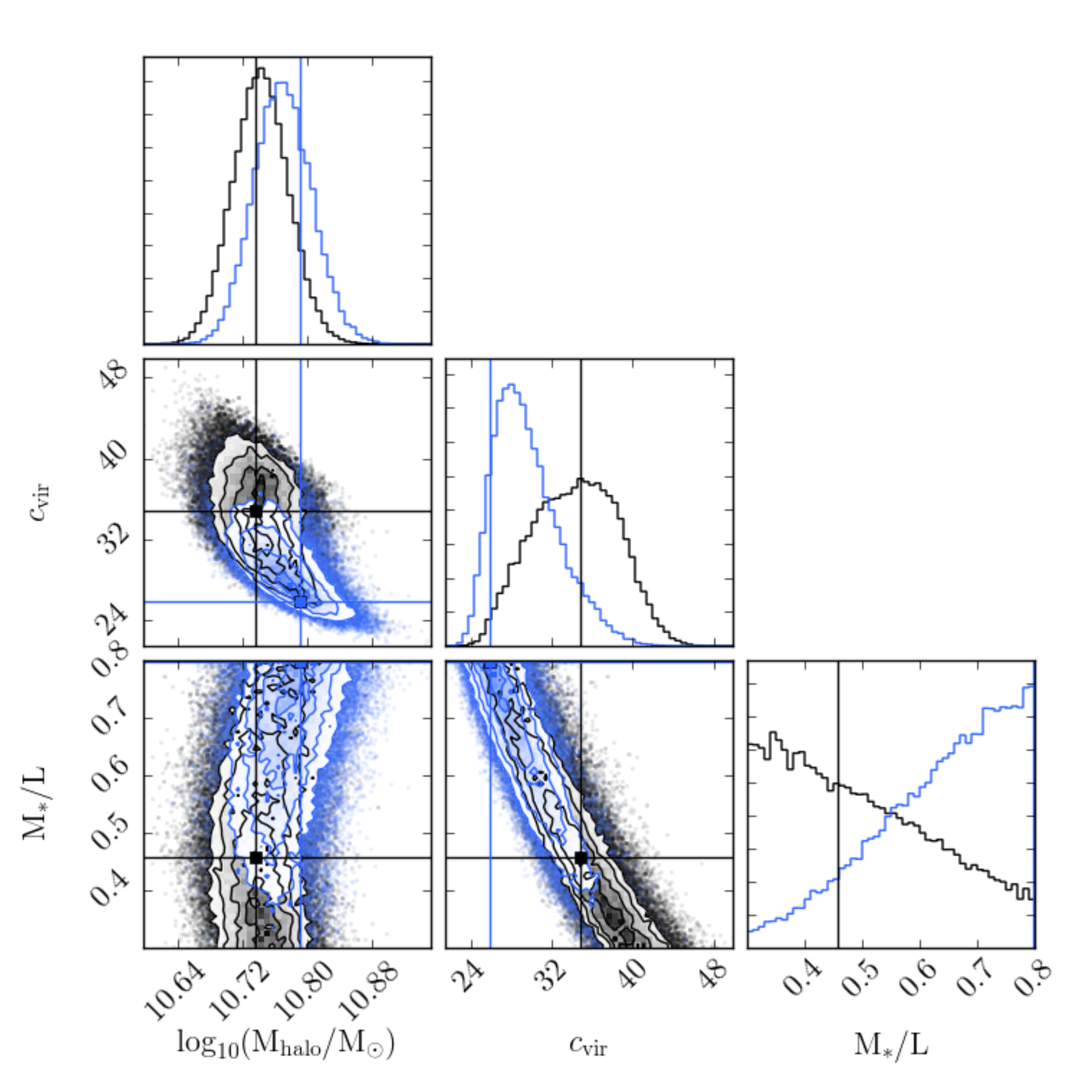}}
\caption{{\bf UGC02259:} See Figure~\ref{DDO161} for caption.}
\label{UGC02259}
\end{figure*}

{\bf UGC02259:} UGC02259 is an example of where the NFW model qualitatively provides a good fit to the data (i.e. has $\chi^2_{\nu}<1.5$) while the DC14 model provides a bad fit to the data  (i.e. has $\chi^2_{\nu}>1.5$).  This likely happens because the inner rising portion of the rotation curve is not well sampled.  Interestingly, this is a lower mass galaxy that is completely DM dominated down to the smallest radius probed.  For the NFW model, the posteriors are very comparable both with and without $\Lambda$CDM priors.  There is not a strong constraint on ${\rm M_*/L}$ although it tends to prefer lower values.  For the DC14 model, we do see that the $\Lambda$CDM priors have a large effect on the fits.  While the mass stays fairly constant, the concentration decreases significantly and the dependence on ${\rm M_*/L}$ switches from preferring low ${\rm M_*/L}$ to high ${\rm M_*/L}$.  The error bars on the middle points in the rotation curve are fairly small which biases the fit to better predict these values.  At these radii, the gas component and disk component switch in terms of which is dominant (see Figure~\ref{UGC02259}), which creates a flattening effect in the rotation curve.  The DC14 model struggles to predict these points, which is why the $\chi^2_{\nu}$ value is so large.  

\begin{figure*}
\centerline{\includegraphics[scale=0.5]{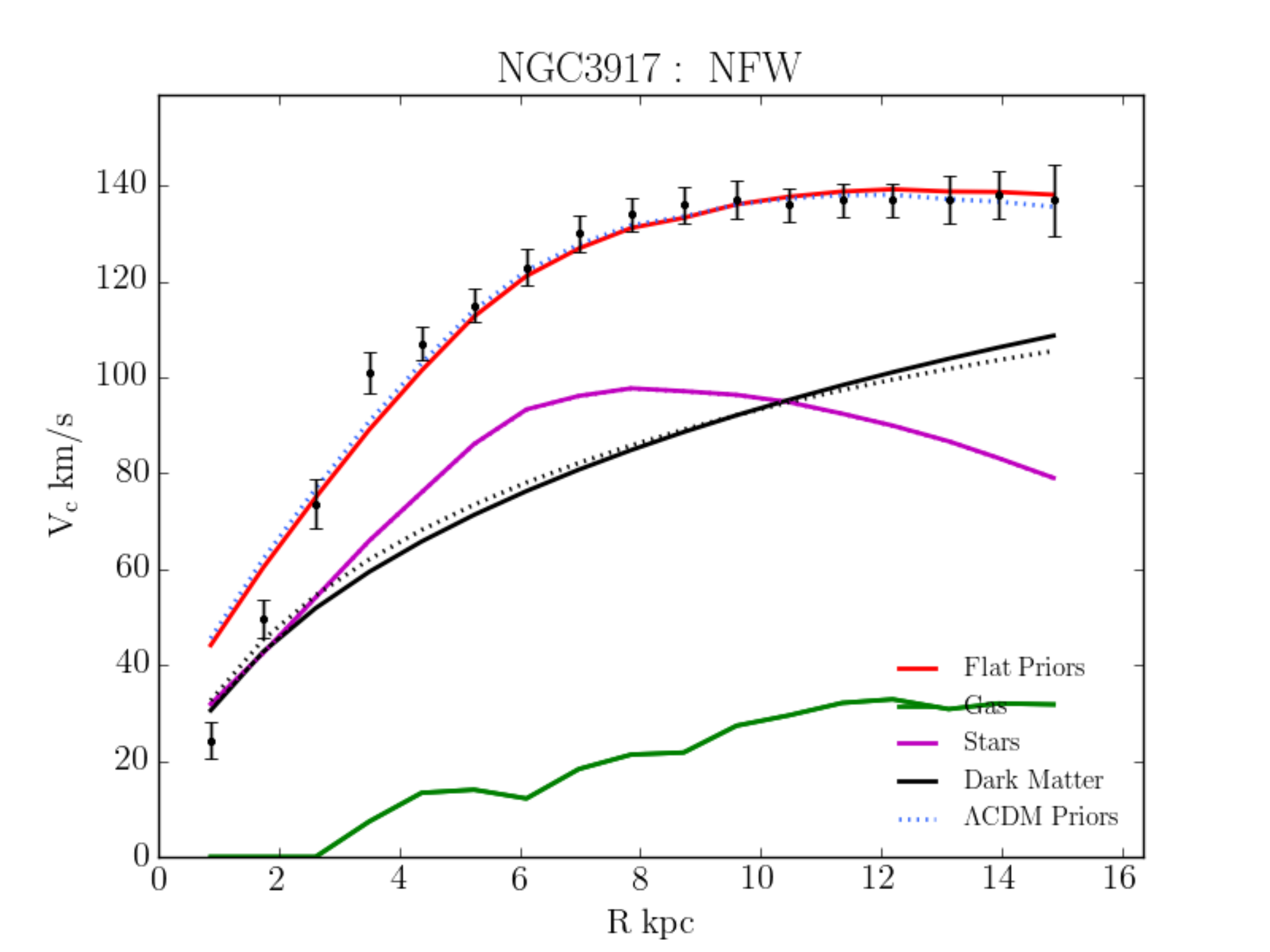}\includegraphics[scale=0.5]{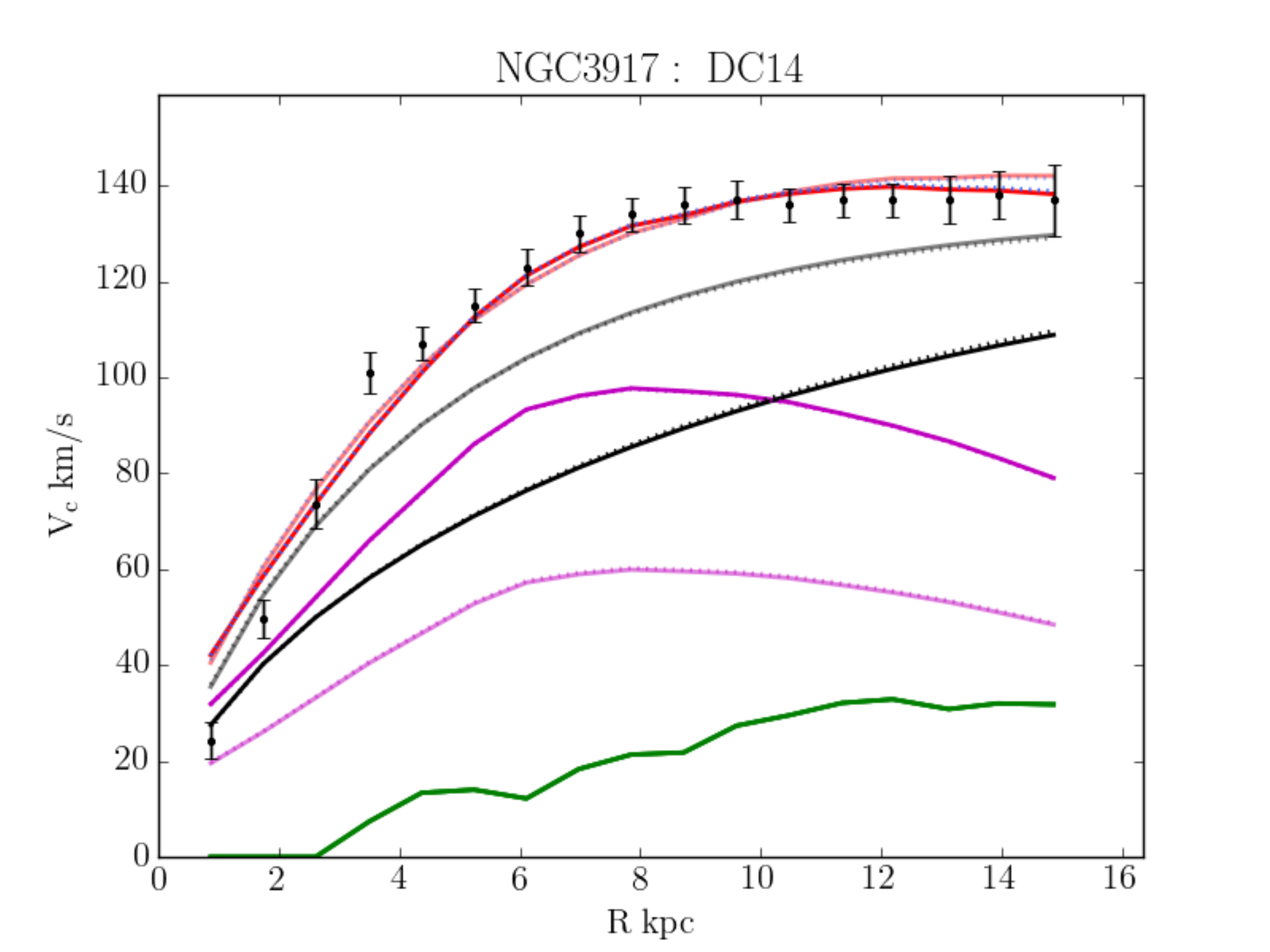}}
\centerline{\includegraphics[scale=0.5]{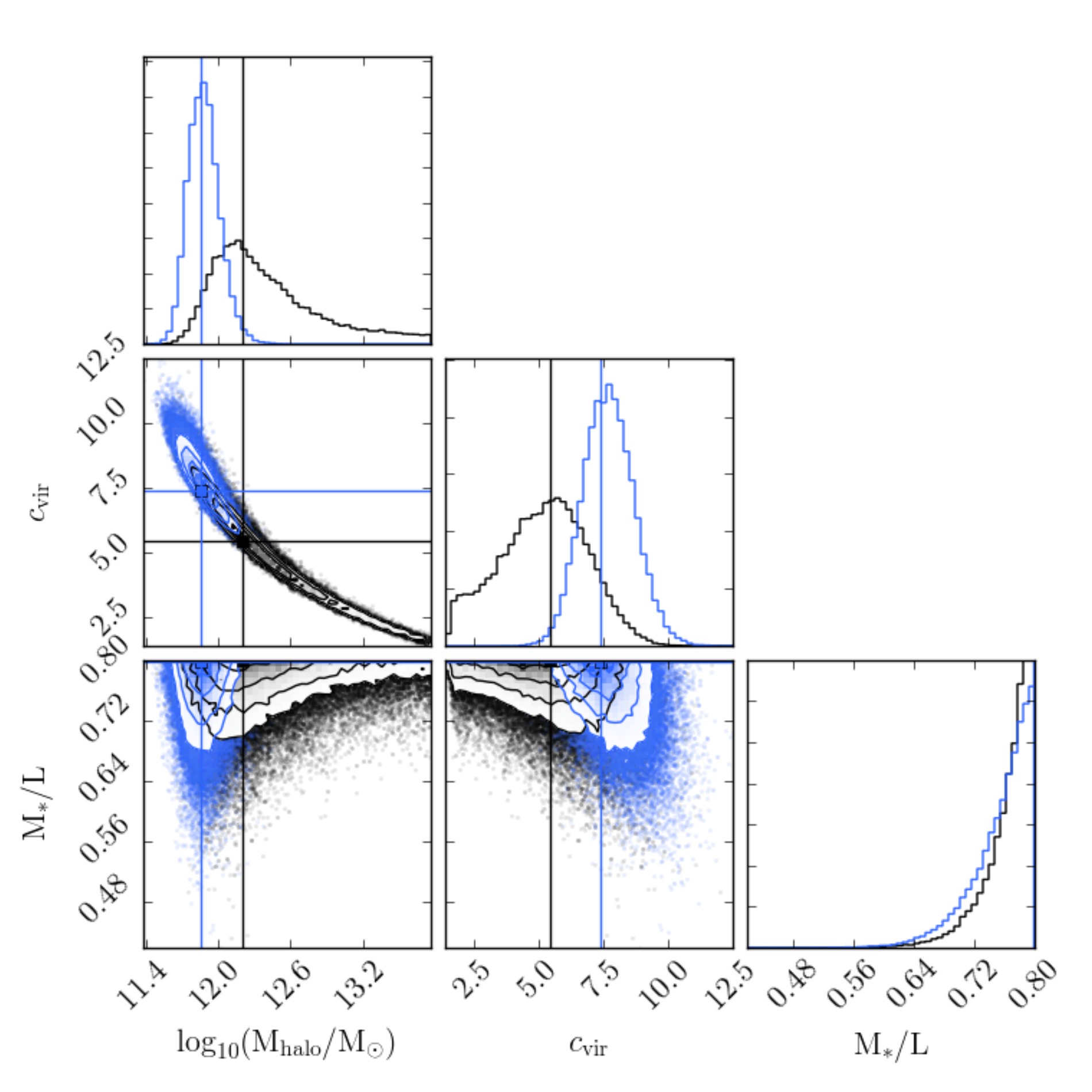}\includegraphics[scale=0.5]{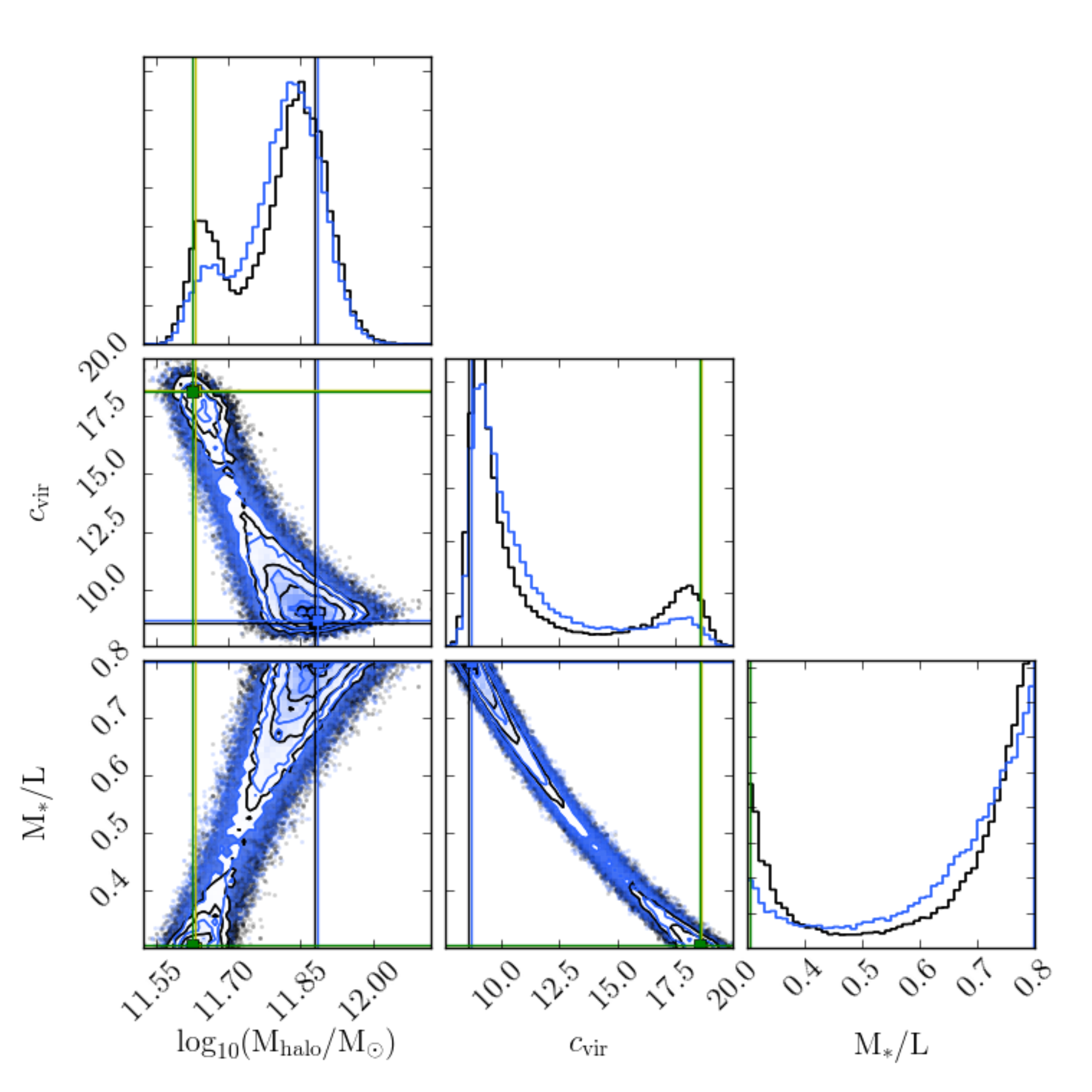}}
\caption{{\bf NGC3917:} See Figure~\ref{DDO161} for caption.}
\label{NGC3917}
\end{figure*}

{\bf NGC3917:}  In Figure~\ref{NGC3917}, we show the posteriors and rotation curve fits for NGC3917.  This case is particularly interesting because it has a very-well-defined multimodal posterior for the DC14 halo model fit.  The galaxy is at the upper end of the mass range where we expect the DC14 model to be valid.  When the $\Lambda$CDM priors are imposed, we can see that the second peak in the distribution shrinks as the first is slightly more preferred although the posterior is still multimodal.  The maximum posterior parameters in each of the peaks do not change when the $\Lambda$CDM priors are imposed so the rotation curve fits for this galaxy are unaffected.  For the NFW model there is clearly only one peak in the data and without $\Lambda$CDM priors we see a large tail in the distribution out to very high-masses.  The 95\% confidence interval on the mass for this galaxy is nearly two orders of magnitude providing very little constraint.  When the $\Lambda$CDM priors are imposed, the posterior parameter space shrinks considerably, although the $\chi^2_{\nu}$ of the maximum posterior halo fit only becomes marginally worse.

For both of these galaxies, none of the fits would be considered ``good" as $\chi^2_{\nu}>3$ for both halo models with and without priors.  ${\rm V_{flat}}$ is well modelled by the fits, however all models struggle in the inner regions where the baryonic contribution is large.

\begin{figure*}
\centerline{\includegraphics[scale=0.5]{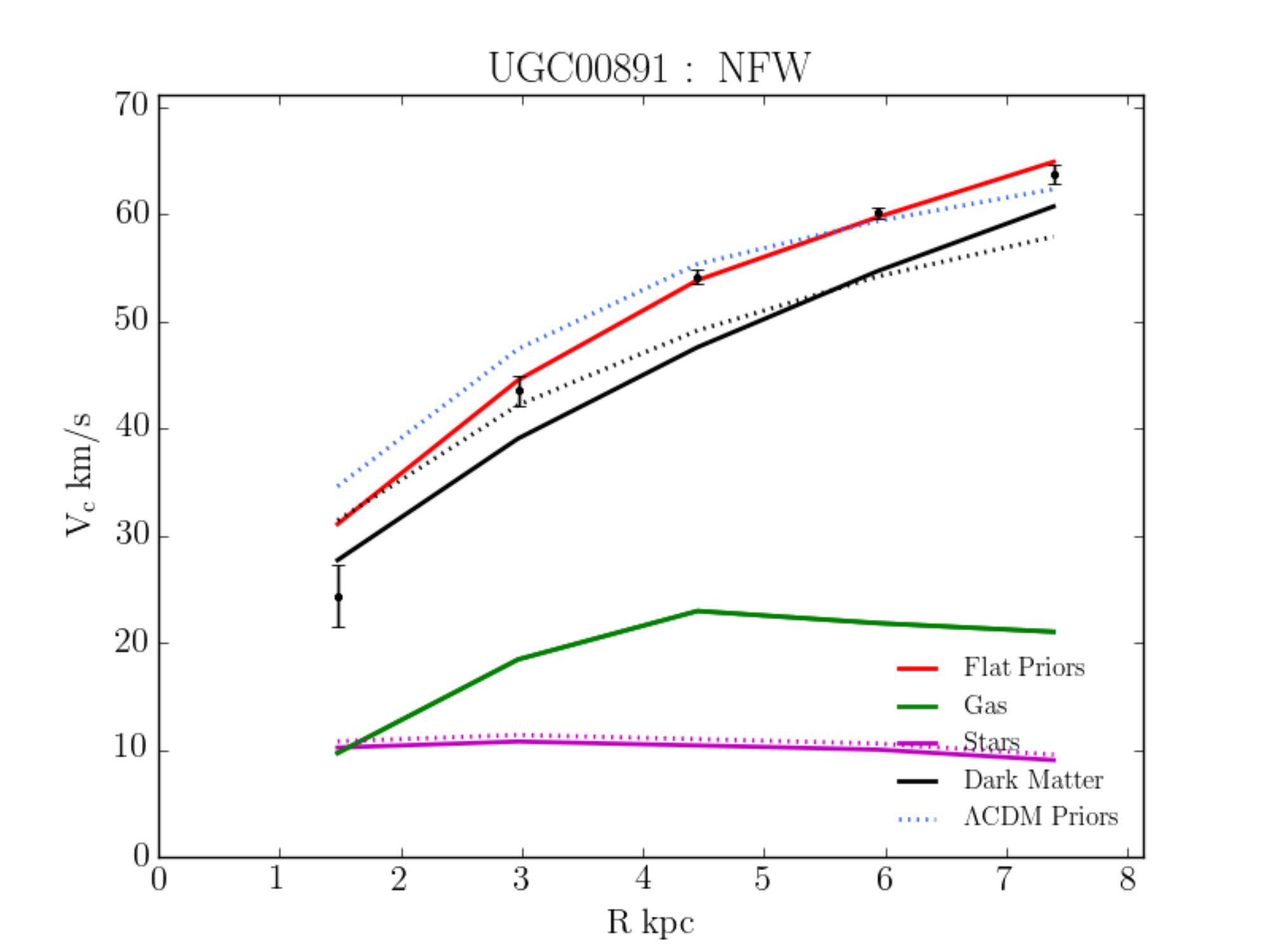}\includegraphics[scale=0.5]{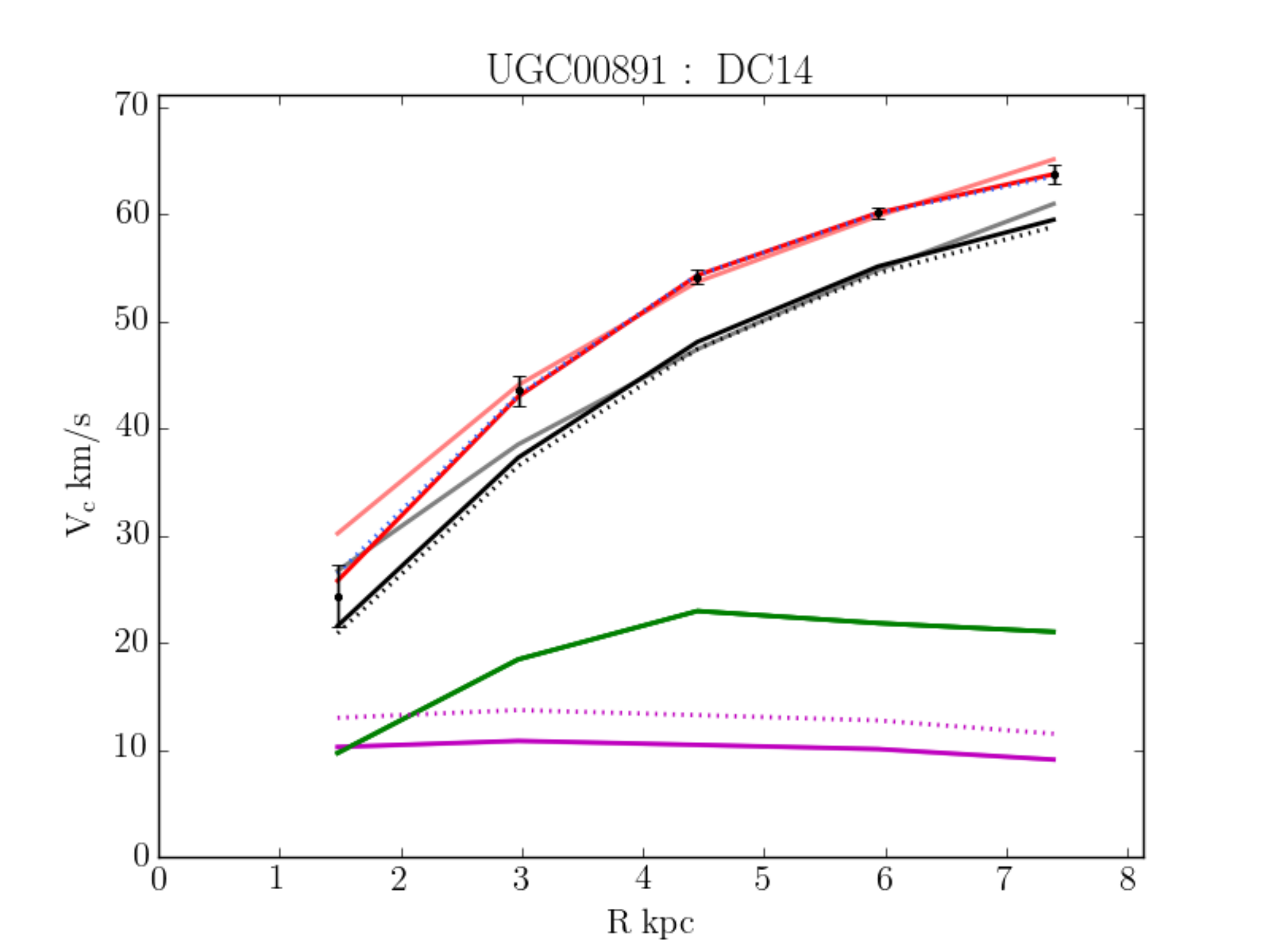}}
\centerline{\includegraphics[scale=0.5]{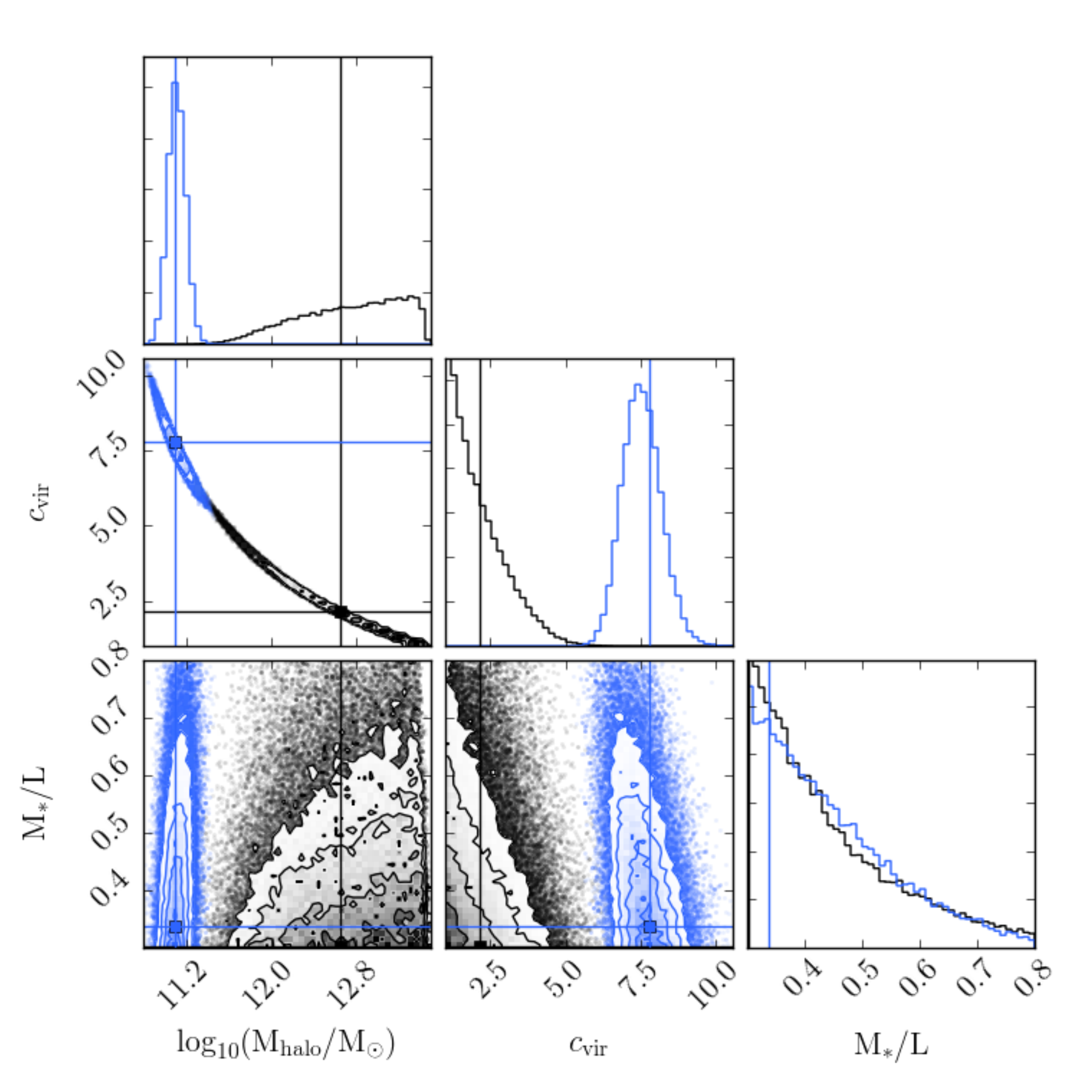}\includegraphics[scale=0.5]{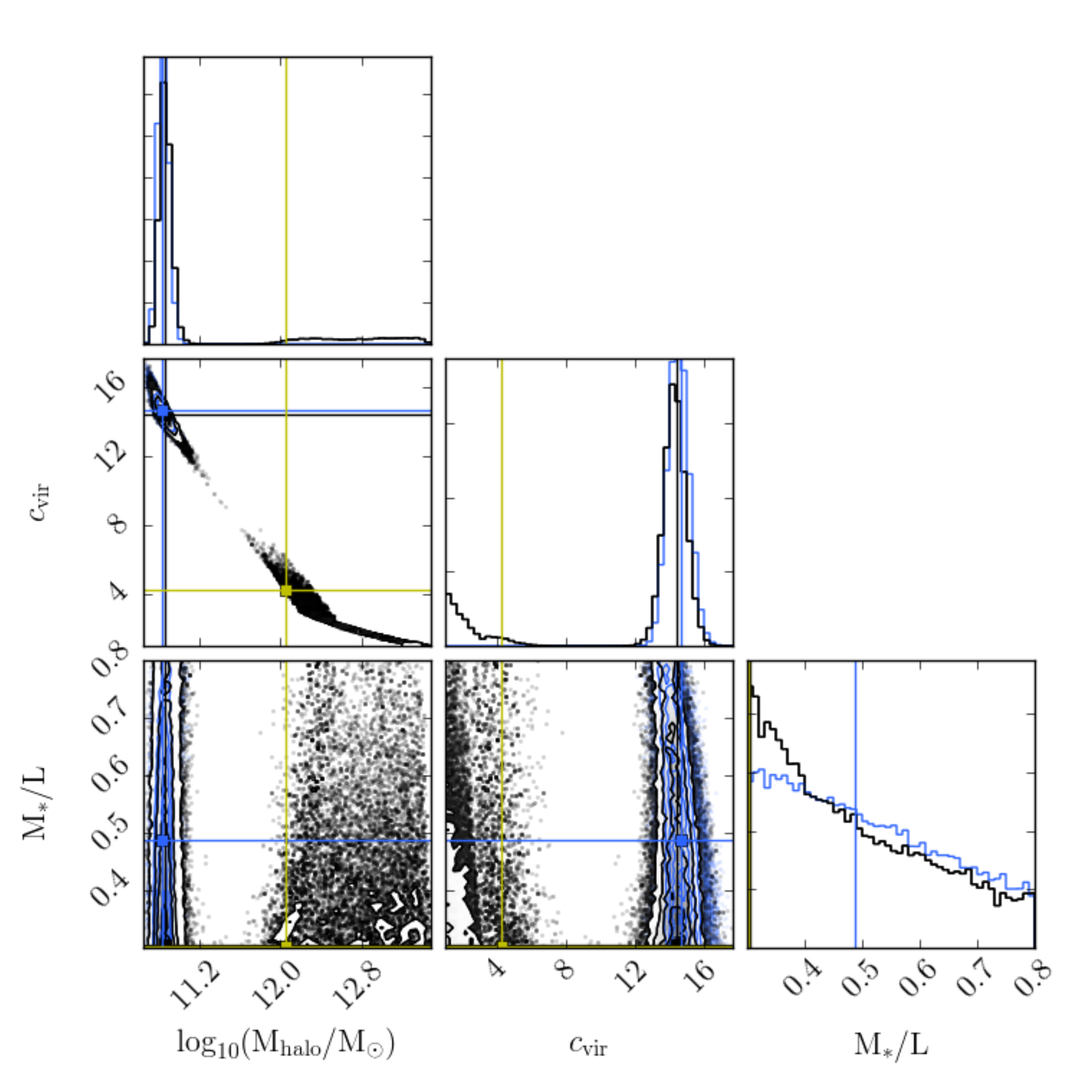}}
\caption{{\bf UGC00891:} See Figure~\ref{DDO161} for caption.}
\label{UGC00891}
\end{figure*}

{\bf UGC00891:} UGC00891 is another example of a galaxy with a multimodal posterior for the DC14 model.  In this case, it has a very strong cosmologically motivated peak as well as a second, much wider mode at very high halo masses (see Figure~\ref{UGC00891}).  When the $\Lambda$CDM priors are imposed, the second mode completely disappears and a very tight constraint is placed on the halo mass and concentration.  For this galaxy, there is no preference on ${\rm M_*/L}$ since the baryonic component is completely gas dominated while the total rotation curve is DM dominated.  Because this is a low-mass, LSB galaxy, it is unsurprising that the NFW model has trouble fitting the rotation curve.  We once again see this characteristic crescent shape in the mass-concentration plane and without $\Lambda$CDM priors, the confidence interval on halo mass spans nearly 1.5 orders of magnitude.  In all three planes, there is barely any overlap between the posterior with the fiducial flat priors and the posterior with the $\Lambda$CDM priors.  The difference in $\chi^2_{\nu}$ between the two halo models is remarkable.  The DC14 model has $\chi^2_{\nu}=0.47(0.74)$ without (with) $\Lambda$CDM priors imposed while for the NFW model $\chi^2_{\nu}=8.17(27.97)$ without (with) $\Lambda$CDM priors imposed. 

\begin{figure*}
\centerline{\includegraphics[scale=0.5]{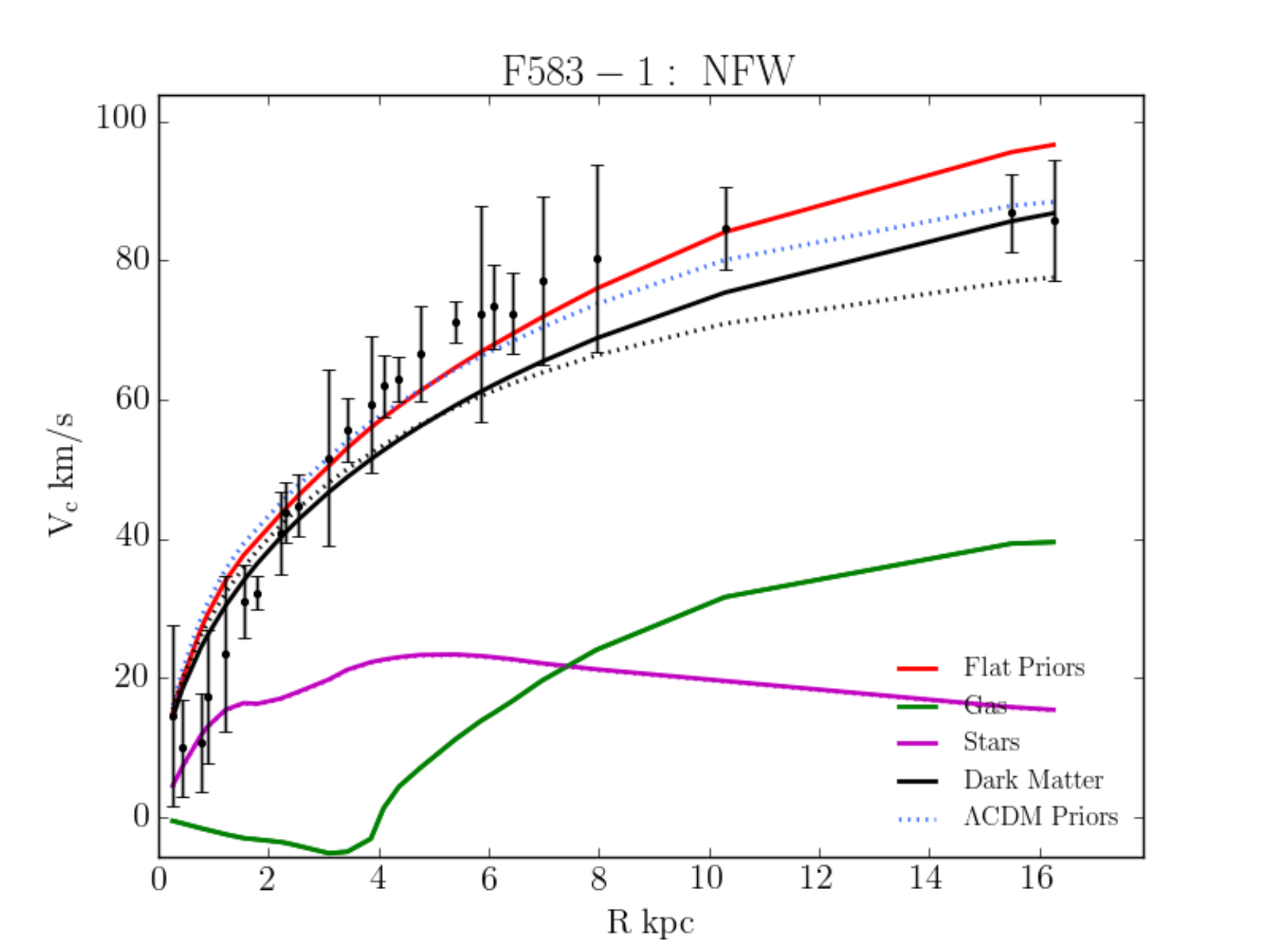}\includegraphics[scale=0.5]{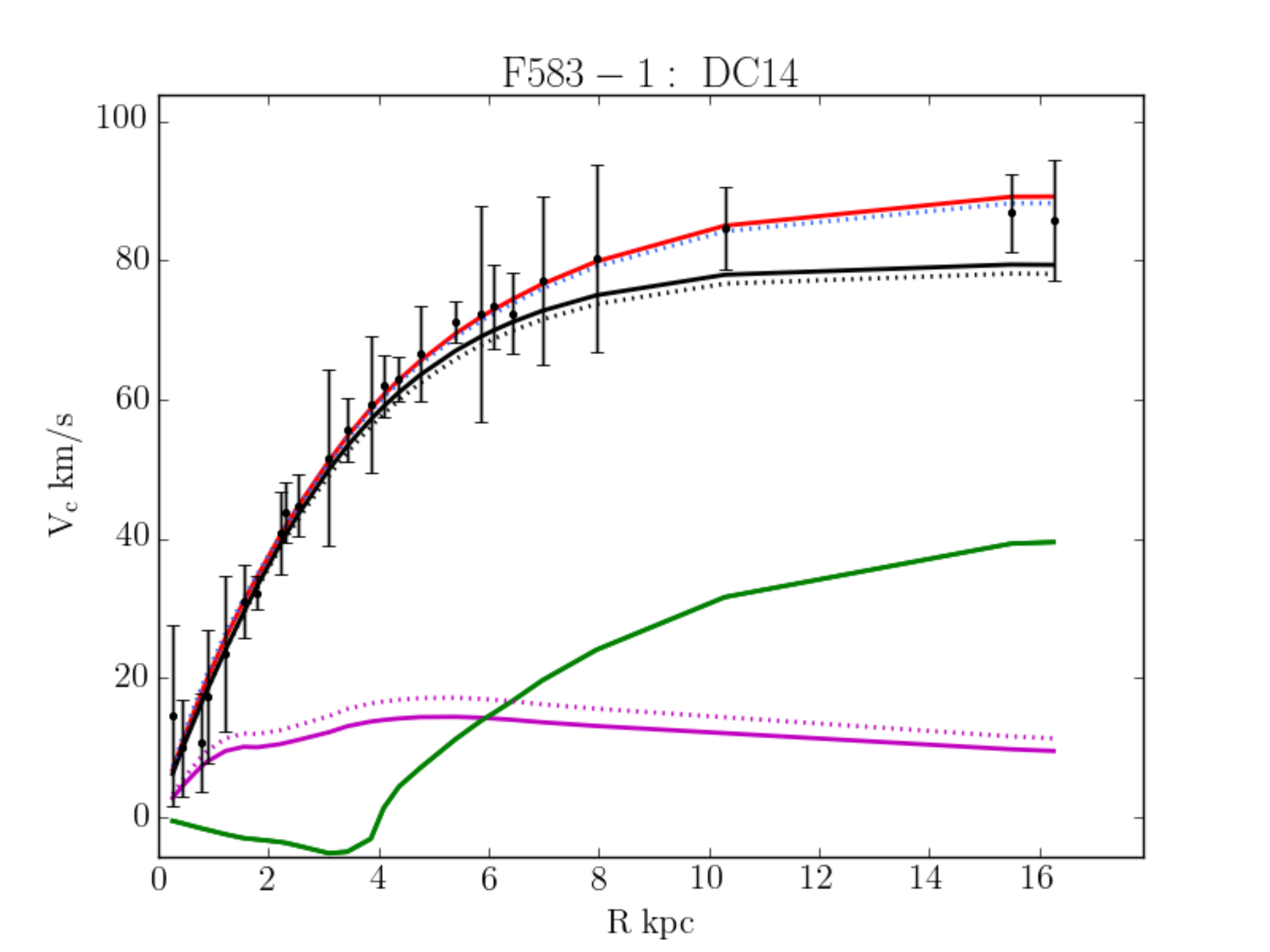}}
\centerline{\includegraphics[scale=0.5]{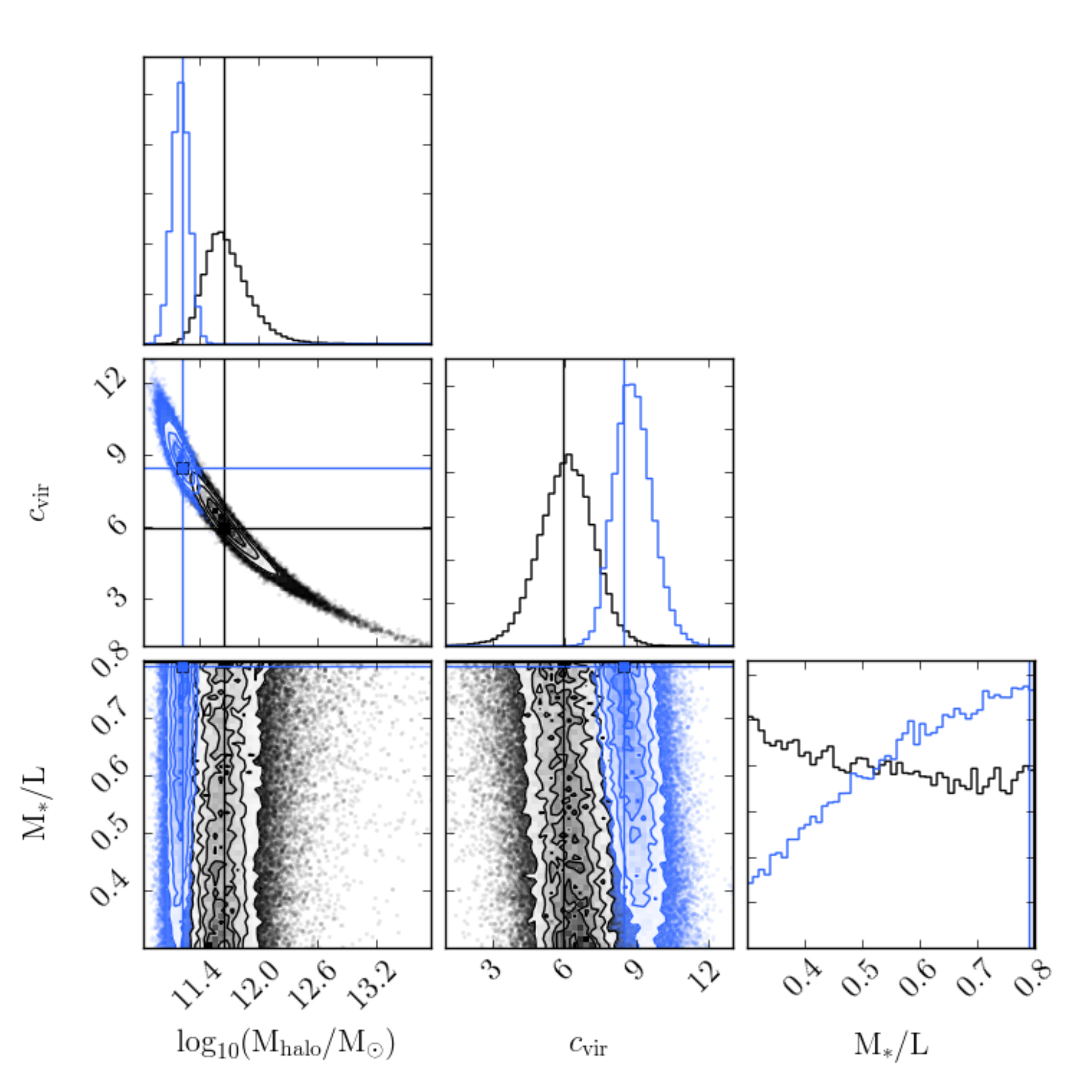}\includegraphics[scale=0.5]{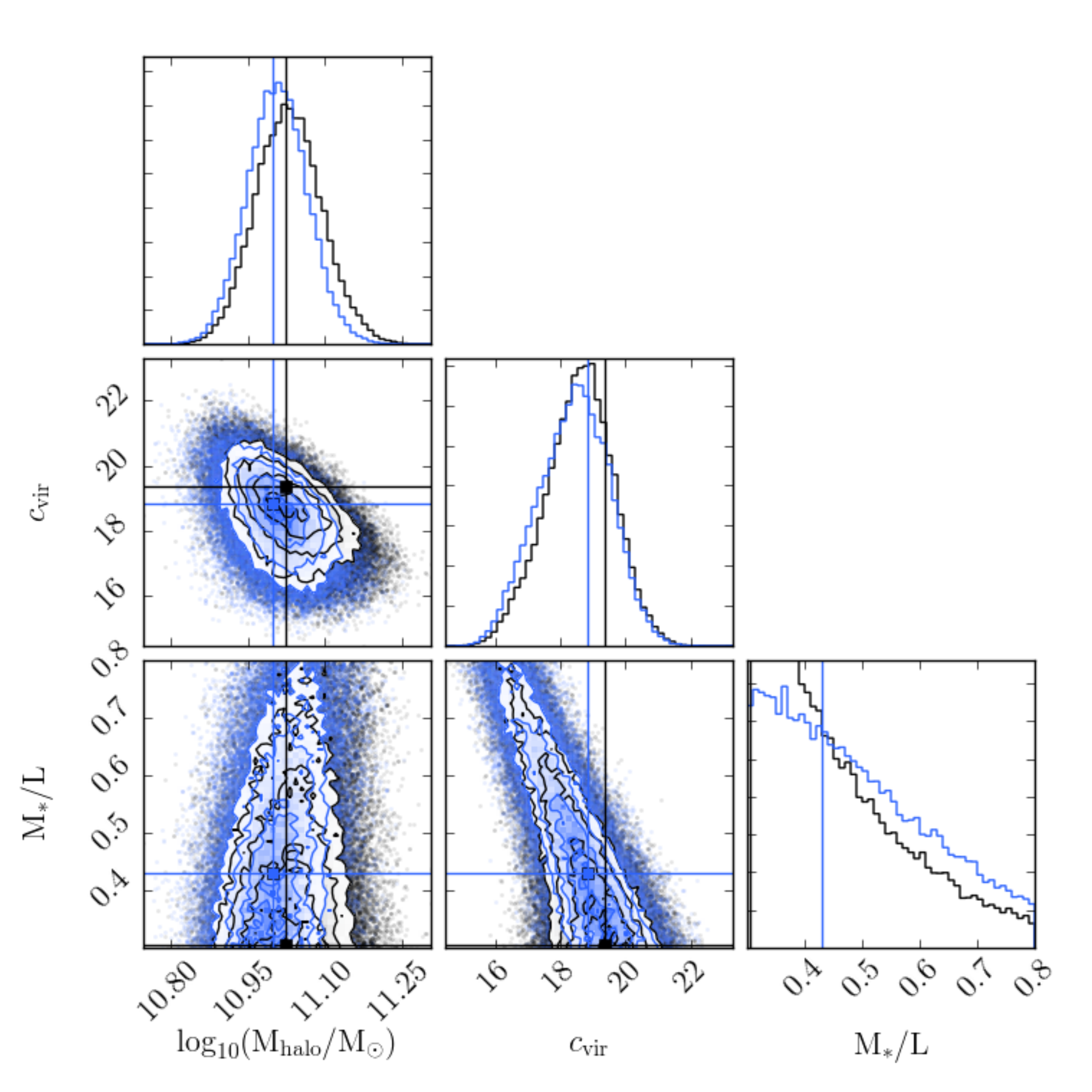}}
\caption{{\bf F583-1:} See Figure~\ref{DDO161} for caption.}
\label{F583-1}
\end{figure*}

{\bf F583-1:} We conclude this section with a galaxy that truly emphasizes the difference between the NFW and DC14 halo models.  In Figure~\ref{F583-1}, we show the posterior distributions and rotation curve fits to this galaxy for the DC14 and NFW models.  It is clear that DC14 does significantly better because this galaxy requires a core.  This is evident in $\chi^2_{\nu}$ which is significantly greater for the NFW model than it is for DC14 ($\chi^2_{\nu}=0.17 (0.21)$ for the DC14 model and $\chi^2_{\nu}=1.69 (1.98)$ for the NFW model without (with) $\Lambda$CDM priors imposed).  We can see that the posterior distributions for the DC14 model nearly lay on top of one another with and without $\Lambda$CDM priors.  This demonstrated that the rotation curve naturally picks out halo parameters for the DC14 model that are consistent with predictions from $\Lambda$CDM.  On the contrary, we see a large change in NFW halo parameters when the $\Lambda$CDM priors are imposed and the posterior distributions before and after are fairly inconsistent.

\section{Scaling with Inner Density Slope}
\label{app1}
\begin{figure*}
\centerline{\includegraphics[scale=0.5]{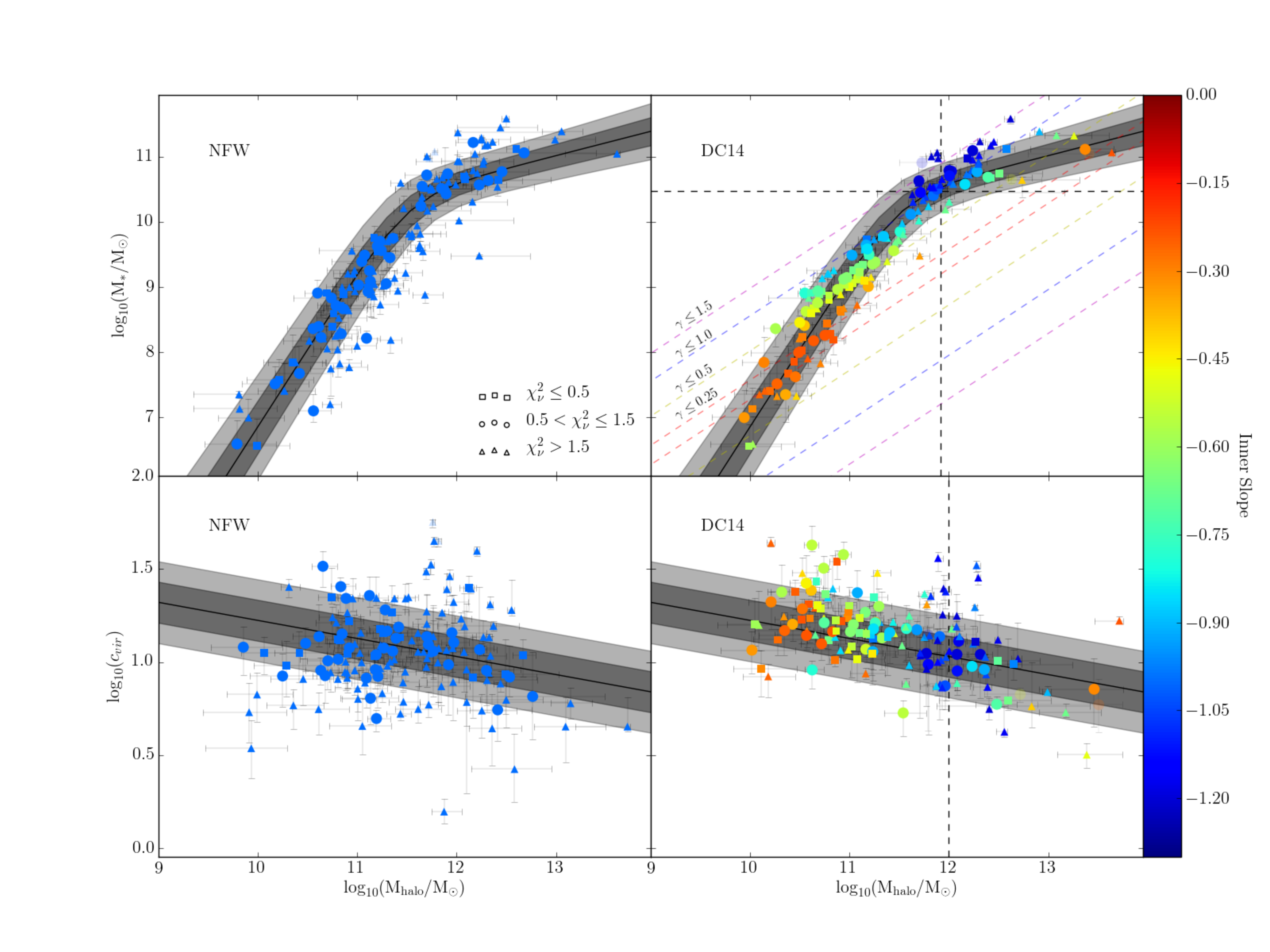}}
\caption{Maximum posterior NFW (left) and DC14 (right) halo fits compared to the abundance matching \protect{\citep{Moster2013}}(top) and mass-concentration relations\protect{\citep{Dutton2014}} (bottom) when the $\Lambda$CDM priors are imposed.  Note that halo mass in the top row represents M$_{200}$ while halo mass in the bottom row represents M$_{vir}$.  The black lines represent the mean relation while the dark and light grey shaded regions show the $1\sigma$ and $2\sigma$ scatter, respetively.  The points are coloured by their inner slope.  The coloured dashed lines in the top-right panel depict lines of constant inner slope for the DC14 model for galaxies of average concentration.  The black vertical dashed lines in the right-hand panels show where the DC14 model is extrapolated outside the range of halo and stellar masses used to predict it.  Error bars represent the projected 95\% confidence interval of the posterior probability distribution.  The lighter, more translucent points in all panels represent the secondary modes for galaxies that have a multimodal posterior.}
\label{lcdminner}
\end{figure*}

In Figure~\ref{lcdminner} we plot the maximum posterior halo parameters against the abundance matching and mass-concentration relations predicted by $\Lambda$CDM, however, here, we have coloured the points by their inner density slopes.  For all NFW haloes, the inner density slope is fixed to a value of $-1$ while this is not the case for DC14.  We see a clear evolution of the inner density slope across the diagrams.  As the mass of the halo increases, the concentration of the haloes decreases and the inner density slope steepens.  

\label{lastpage}
\end{document}